\documentclass{gAPA2e}

\usepackage[pdftex, citecolor=black, urlcolor=black,
  linkcolor=black, colorlinks=true]{hyperref}

\newcommand\ep{\varepsilon}

\newcommand\norm[2]{\left\Vert #1 \right\Vert_{#2}}
\newcommand\twoscale{\overset{_2}{\rightharpoonup}}
\newcommand\strongtwoscale{\overset{_2}{\rightarrow}}

\newcommand\pd[2]{\frac{\partial{#1}}{\partial{#2}}}

\newcommand{\seq}[1]{\left\{{#1}\right\}}

\newcommand\mv[1]{\langle #1 \rangle}

\newcommand{\spec}[1]{\sigma{\left( #1 \right)}}

\newcommand\dblint[2]{\int_{#1}{\int_{Q}{#2}  \ \mathrm{d}y} \mathrm{d}x}

\title{Homogenisation and spectral convergence of a periodic elastic composite with weakly compressible inclusions.}
\author{Shane Cooper\thanks{Email address: shane.cooper@fresnel.fr} \\ Institut Fresnel, Aix-Marseille Universit\'{e} (UMR CNRS 7249), Domaine Universitaire de Saint-J\'{e}r\^{o}me F13397, Marseille cedex 20, France} \received{\today}

\begin{document}

\maketitle
\begin{abstract}
A two phase elastic composite with weakly compressible elastic inclusions is considered. The homogenised two-scale limit problem is found, via a version of the method of two-scale convergence, and analysed. The microscopic part of the two-scale limit is found to solve a Stokes type problem and shown to have no microscopic oscillations when the composite is subjected to body forces that are microscopically irrotational.  The composites spectrum is analysed and shown to converge, in an appropriate sense, to the spectrum of the two-scale limit problem. A characterisation of the two-scale limit spectrum is given in terms of the limit macroscopic and microscopic behaviours.
\end{abstract}

\section{Introduction}
It is well known that composite materials often display physical properties that are not observed by their individual constitutive parts. 
This leads to the question whether one can produce composite materials to exhibit a desired property. Stated differently, can one determine the effective properties of a composite material with prescribed microscopic data. Mathematically, one can approach this question via the application of homogenisation theory to determine the `homogenised' limit to the equations modelling the composite'	s behaviour. Such limit equations can be then considered to contain the effective physical properties of the composite with the limit solutions describing the effective behaviour. Historically, ``Classical'' homogenisation has been applied to composites with moderately contrasting heterogeneity. Here the heterogeneity is replaced by an equivalent homogeneous medium with uniform physical properties, and is therefore incapable of describing a range of interesting and unusual effects. Subsequently, homogenisation was used to study composite materials with highly contrasting coefficients. The so-called high contrast homogenisation theory has been used to describe many non-trivial and interesting behaviours; examples include memory effects (e.g. \cite{FeKru80,khrus, SAN})  and other non-local effects (e.g. \cite{ChenSmZh,chen1,bell,avila}). 

A useful analytical tool in the homogenisation theory is the method of two-scale convergence first introduced by Nguesteng \cite{nguetseng} and substantially developed by Allaire \cite{allaire} particularly in the context of high contrast periodic problems. Two-scale convergence was further developed by Zhikov for the study of high-contrast spectral problems in bounded \cite{zhikov1} and unbounded \cite{zhikov2} domains. Therein, Zhikov described a ``two-scale limit operator" and the Hausdorff convergence of spectra in terms of strong two-scale resolvent convergence and the two-scale compactness of eigenfunctions. Zhikov also explicitly described the ``limiting" spectrum by a coupled system of limit equations in terms of the macroscopic and microscopic variables. Furthermore, Zhikov showed upon decoupling, the effective macroscopic properties to depend non-linearly on the spectral parameter, essentially giving rise to a description of a ``microresonace'' effect: a distinct change in physical properties when the applied `macroscopic' frequency is close to the eigenfrequencies of the microscopic inclusions. In the context of electromagnetism these results first due to Zhikov, \cite{zhikov1}, where interpreted as the appearance of effective negative magnetism for appropriately polarised waves of certain frequencies in a non-magnetic material with high contrasting electric permeability, see \cite{BouchFel}.

Recently, the interest in high-contrast homogenisation has been boosted by applications to so-called metamaterials, c.f. \cite{pen}. This class of composite materials exhibit macroscopic or `effective' physical properties not commonly found in nature, such as electromagnetic meta-materials with negative refractive index, cloaking devices and super lens, to name but a few. More recently, Smyshlyaev following \cite{ChenSmZh}, showed in \cite{VPS}, using multi-scale asymptotic expansions, that composite materials with a ``partially high contrast'' between constitutive parts are capable of accounting for additional phenomena such as directional propagation. Thus one may find in the study of this broader class of composite materials the physical effects one seeks in meta-materials. In \cite{ivkvps} the two-scale homogenisation analytic tools are developed for a broad class of such problems. Here, we shall use some of these tools to study one particular problem of this class that arises in elasticity.

In this paper we study a partially high contrast elastic composite whose `inclusion' phase is disjoint and periodically distributed through the `matrix' phase. The matrix phase is an arbitrary, generally heterogeneous material with a uniformly positive elasticity tensor and the inclusion phase is considered to be isotropic and `soft' in shear. Namely, the shear modulus for the inclusion material is chosen to be of the order $\ep^2$, where $\ep$ is the composite's periodicity size, while the bulk modulus remains uniformly positive. Such elastic inclusions can be called weakly compressible and elastic composites that are weakly compressible everywhere were studied in \cite{BakhEgl}. Composites with weakly compressible inclusions were first studied by Panasenko, via the method of asymptotic expansions, for $\ep$-independent body forces, c.f. \cite{pan2,pan3}. Therein, Panasenko showed the homogenised limit equations to be a system of equations depending only on the macroscopic spatial variable, i.e. not a two-scale system. A novelty of our work is that the body forces are allowed to be $\ep$-dependent, thus eventually allowing for a study of the corresponding spectral problem. The resulting homogenised limit equations are found to be genuinely two-scale, see Theorem \ref{thm:elastprob2}, which reduce to the classical case for not only macroscopic body forces, as shown in \cite{pan2,pan3}, but for a broader class of  applied body forces, namely microscopically irrotational forces, see Corollary \ref{cor:elastprob1}. We perform spectral analysis and show that the spectrum corresponding to the original problem converges in the sense of Hausdorff to the spectrum of the two-scale limit problem, see Theorem \ref{spectralcompactness}. Furthermore, due to the behaviour of the limit eigenfunctions prescribed by the weakly compressible condition, the two-scale limit spectral problem appears uncoupled in the macroscopic and microscopic problems. In the course of obtaining these results, we apply and develop further an appropriate modification of the two-scale convergence techniques to deal with the partial degeneracy, cf \cite{ivkvps}.

The layout of the paper is as follows: Section \ref{sec:probform} contains the problem  formulation and presentation of the main results. Section \ref{background} reviews the necessary background material, in particular the method of two-scale convergence, its applications to homogenisation and further modifications to deal with partial degeneracies. Section \ref{sec:hom} and \ref{sec:spectral} are dedicated to the proofs of the main homogenisation theorem and the spectral convergence respectively.

\section{Problem formulation and main results}
\label{sec:probform}
\begin{figure}[htb]
	\centering
		\includegraphics[width=1.00\textwidth]{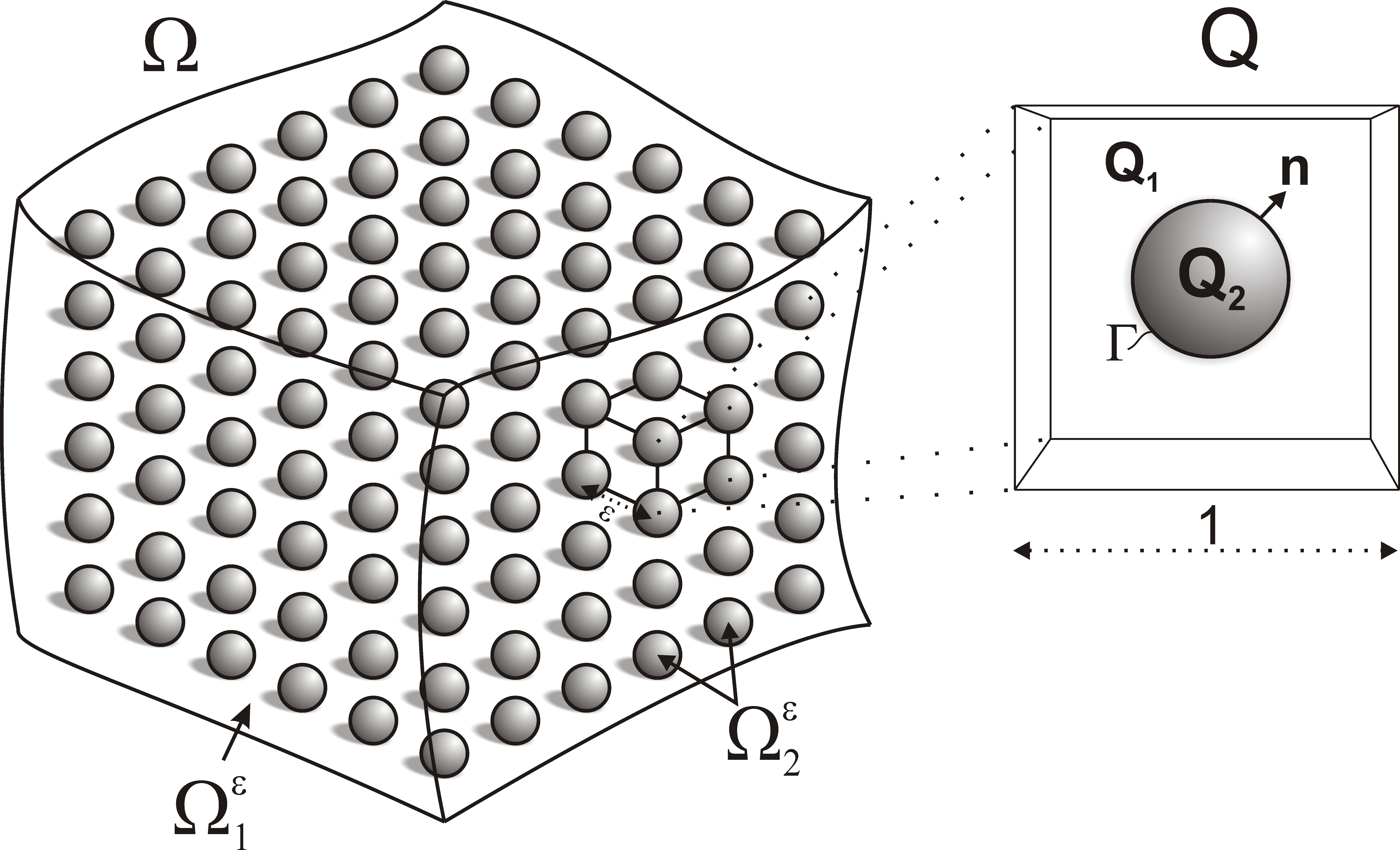}
		\caption{An example of a periodic composite body considered in this paper.}
	\label{fig:composite}
\end{figure}
We begin by introducing the notation used throughout this paper. $\Omega \subset \mathbb{R}^d$, $d \ge 2$, open and bounded, shall denote the domain occupied by the composite material, see Figure \ref{fig:composite}. $Q = [0,1)^d$ is the periodic reference cell and shall consist of two disjoint regions: the `inclusion' $Q_2$, a subset of $Q$ with smooth boundary $\Gamma$ and the `matrix' $Q_1 = Q \backslash \overline{Q_2}$. We assume that $Q_1$ is connected and $Q_2$ does not intersect the boundary $\partial{Q}$, i.e. $\overline{Q_2} \subset (0,1)^d$. We denote by $F_2$ the $Q$-periodic extension of $Q_2$ throughout $\mathbb{R}^d$ and by $\ep F_2$ the $\ep$ contraction of $F_2$, i.e.
$$
F_2 : = \left\{ x : \text{ $x = y + k$ for some $y \in Q_2$ and some $k \in \mathbb{Z}^d$}  \right\},
$$
and $\ep F_2 = \left\{ x : x / \ep \in F_2 \right\}$. We denote by $\Omega^\ep_2$ and $\Omega^\ep_1$ the inclusion phase and matrix phase respectively. That is, $\Omega^{\ep}_2 = \Omega \cap \ep F_2$ and  $\Omega^{\ep}_1 = \Omega \backslash \overline{\Omega^{\ep}_2}$. For a Banach space $X$ the spaces $[X]^d$ shall denote the space of vector-valued functions whose components belong to X, i.e. for $ u \in [X]^d$, $u = \left\{ u_1, u_2, \ldots, u_d \right\}$ and $u_1, u_2, \ldots, u_d \in X$. The spaces of matrix-valued functions $[X]^{d \times d}$ and fourth rank tensor-valued functions $[X]^{d \times d \times d \times d}$ shall be understood in a similar way.

We shall consider the following problem:
\begin{equation}
\label{resolventpde}
 \begin{split}
- \nabla \cdot \Big( C^{\ep} e( u )\Big) + \alpha u & =  f^{\ep}  \quad \text{in $\Omega$} \\
u & = 0 \quad \text{ on $\partial{\Omega}$}.
\end{split} 
\end{equation}
$\alpha \ge 0$. The case $\alpha = 0$ is the elastostatic problem with body force $f^\ep$ while $\alpha > 0$ is the resolvent problem, which is important for spectral analysis (cf. \cite{zhikov2}.) The underlying density function $\rho(y)$ could be taken to be any positive matrix but is, for simplicity of deviation, assumed to be equal to unity. Here $u \in [H^1_0(\Omega)]^d$ is the unknown displacement and $e(u)_{ij}=\frac{1}{2}\left( \pd{u_i}{x_j} + \pd{u_j}{x_i} \right)$ is the infinitesimal strain tensor. Furthermore, $C^{\ep}(x) = C(x/\ep)$ for $C(y)\in [L^{\infty}(Q)]^{d \times d \times d \times d}$, is the elasticity tensor of the composite material which is considered to be  $Q$-periodic, positive definite in the matrix and isotropic in the inclusion with Lam\'{e} coefficients $\lambda \sim O(1)$, $\mu \sim O(\ep^2)$. Explicitly $C(y)$ is of the form:
\begin{equation}
\label{elast:degentensor}
\begin{split}
 C(y) = C^{(1)}(y) &+ \ep^2C^{(0)}(y), \\
C^{(1)}_{ijpq}(y)  = \chi_1(y)C^{(2)}_{ijpq}(y)  + \chi_2(y)\delta_{ij}\delta_{pq}, & \ C^{(0)}_{ijpq}(y) = \chi_2(y)( \delta_{ip}\delta_{jq}  + \delta_{iq}\delta_{jp}).
\end{split}
\end{equation}
(We have set, for simplicity, $\lambda = 1$, $\mu = \tfrac{\ep^2}{2}$.) Here $\delta$ is the Kronecker delta symbol, $\chi_i$ is the characteristic function of $Q_i$, $C^{(2)} \in [L^{\infty}(Q)]^{d \times d \times d \times d}$ is taken to be symmetric and positive definite: 
\begin{align}
\label{sympos}
C^{(2)}_{ijpq}(y) &= C^{(2)}_{jipq}(y) = C^{(2)}_{pqij}(y), & C^{(2)}_{ijpq}(y)\eta_{pq}\eta_{ij} & \ge \nu \vert \eta \vert^2,
\end{align}
for some $\nu >0$, for all $y \in Q_1$, for all symmetric $\eta$; $f^{\ep}(x) \in [L^2(\Omega)]^d$ is a prescribed externally applied `body force'. \par 
 We can see from \eqref{elast:degentensor}-\eqref{sympos} that $C^{(1)}$ is symmetric and non-negative while $C^{(1)} + C^{(0)}$ is symmetric and positive definite, i.e.
\begin{gather}
\label{eq:elasthom2}
\begin{aligned}C^{(l)}_{ijpq}(y) &= C^{(l)}_{jipq}(y) = C^{(l)}_{pqij}(y), & \text{for $l=0,1.$} \end{aligned} \\ \nonumber
\begin{aligned}C^{(1)}_{ijpq}(y)\eta_{pq}\eta_{ij}  &  \ge 0, &  \left(C^{(1)}_{ijpq}(y) + C^{(0)}_{ijpq}(y)\right)\eta_{pq}\eta_{ij}    \ge \nu \vert \eta \vert^2, \end{aligned}
\end{gather}
for some $\nu >0$, for all $y \in Q$, for all symmetric $\eta$. In \eqref{eq:elasthom2} and henceforth summation is implied with respect to repeated indices. We now state our first main result.

\begin{theorem}
\label{thm:elastprob2}
Let $f^\ep(x)$ weakly (strongly) two-scale converge to $f(x,y)$ as $\ep \rightarrow 0$. Then the sequence $u^\ep$, of solutions to \eqref{resolventpde}, weakly (strongly) two-scale converges to $u^0(x,y) = u(x) + v(x,y)$ as $\ep \rightarrow 0$, where $(u,v) \in [H^1_0(\Omega)]^d \times [L^2(\Omega ;H^1_{0}(Q_2))]^d$ is the unique solution to
\begin{gather}
\label{eq:elastprob2} 
- \nabla \cdot \left( C^{\mathrm{hom}} \nabla u(x) \right)   + \alpha u(x) + \alpha\mv{v}(x) = \mv{f}(x)  \quad \text{in $\Omega$}, \\ \label{eq:elastprob3}
\begin{aligned}
 - \Delta_y v(x,y) + \alpha v(x,y) + \alpha u(x) & = f(x,y) + \nabla_y p(x,y) \quad \text{in $Q_2$} \\
 \nabla_y \cdot v(x,y) & = 0 \quad \text{in $Q_2$}   \\
 v(x,y) & = 0 \quad \text{on $\partial{Q_2}$},
\end{aligned}
\end{gather}
where $p \in L^2(\Omega ; H^1(Q_2))$ is unknown. $\mv{\cdot}$ denotes the mean value over $Q$, i.e. $\mv{f}(x) = \int_Q f(x,y) \ \mathrm{d}y$. $C^{\mathrm{hom}}$ is the constant coefficient positive definite tensor given by
\\
\begin{equation}
\label{elastinserted1}
C^{\mathrm{hom}}_{ijrs} = \int_{Q} C^{(1)}_{ijpq}(y) \left( \delta_{pr}\delta_{qs} + \pd{N^p_{rs}}{y_q} \right) \mathrm{d}y.
\end{equation}
Here $N_{rs} = (N^1_{rs},N^2_{rs},\ldots,N^d_{rs})$ is a $Q$-periodic solution to the degenerate cell problem
\begin{equation}
\label{eq:ehom0.2}
- \nabla_y\cdot \left(C^{(1)}(y) \left( e_r \otimes e_s + \nabla_y N_{rs}(y) \right) \right)  = 0  \quad \text{in $Q$}.
\end{equation}
\end{theorem}
The solution $N_{rs}$ to the degenerate cell problem \eqref{eq:ehom0.2} is not unique as non-trivial solutions to the homogeneous problem exist due to the degeneracy of $C^{(1)}$. Such homogeneous solutions form a non-trivial subspace of $[H^1_{\#}(Q)]^d$ and are characterised by satisfying the condition $C^{(1)}(y)\nabla_y v = 0$ in $Q$. However,  we can see from \eqref{elastinserted1} that $C^{\text{hom}}$ remains unchanged if we add any homogeneous solution of the degenerate cell problem to $N$; from this we can conclude that even though the solutions to the degenerate cell problem are not unique, $C^{\text{hom}}$ is unique. Furthermore $C^{\text{hom}}$ is positive definite, which can be seen by noticing from the variational form characterisation of the tensor  $C^{\text{hom}}$ that it is ``more positive" than the effective tensor corresponding to the classical homogenisation of a perforated domain problem, which is indeed positive, see e.g. \cite{CPS}. 

Let us now consider problem \eqref{eq:elastprob3}. One notices that for microscopically irrotational body forces, namely forces of the form $f(x,y) = f_0(x) + \nabla_y f_1(x,y)$, one can choose $p$ in \eqref{eq:elastprob3} in such a way as to `absorb' the forcing term, i.e. by setting $p(x,y) = - y\cdot \big( f_0(x) - \alpha u(x) \big) - f_1(x,y) + \tilde{p}(x,y)$. As a result, \eqref{eq:elastprob3} is reduced to the homogeneous Stokes problem: find $v, \tilde{p},$ such that
\begin{align*}
 - \Delta_y v(x,y) + \alpha v(x,y)  & = \nabla_y \tilde{p}(x,y) \quad \text{in $Q_2$} \\
 \nabla_y \cdot v(x,y) & = 0 \quad \text{in $Q_2$}   \\
 v(x,y) & = 0 \quad \text{on $\partial{Q_2}$},
\end{align*}
which is well known to have only the trivial solution $v \equiv 0$, see e.g. \cite{temam}. This shows that for microscopically irrotational body forces  the limit problem \eqref{eq:ehom0.1} is independent of the microscopic variable $y$, i.e. the limit solution $u^0(x,y) = u(x)$ has no microscopic oscillations. Furthermore, in this case $u^\ep \rightarrow u$ strongly in $L^2$ when $f^\ep(x) \strongtwoscale f(x,y)$. These striking results are very different to the double porosity case where that the homogenised limit is of a genuine two-scale nature, for a general body force $f^\ep$, with the limit function depending on both the macroscopic and microscopic variable. The reason for the limit function having no microscopic oscillations, for the described body force, is precisely due to the form of the partial degeneracy: the inclusion phase is isotropic with Lam\'{e} coefficient $\lambda \sim 0(1)$, which means that in the asymptotic limit $\ep \rightarrow 0$ the inclusion phase is microscopically incompressible, i.e. $\nabla_y \cdot v =0$, and, as a consequence, for microscopically  irrotational body forces no deformations will occur on the microscopic scale. The following corollary states these observations in a precise form.

\begin{corollary}
\label{cor:elastprob1}
Let $f^\ep(x)$ strongly two-scale converge to $f(x,y) = f_0(x) + \nabla_y f_1 (x,y)$ for given, sufficiently regular, $f_0,f_1$. Then the sequence $u^\ep$, of solutions to \eqref{resolventpde}, strongly converges to $u(x)$ in $L^2(\Omega)$ as $\ep \rightarrow 0$, where $u\in [H^1_0]^d$ is the unique solution to
\begin{equation} 
\label{eq:ehom0.1}
-  \nabla \cdot \left( C^{\mathrm{hom}} \nabla u(x) \right)   + \alpha u(x) = \mv{f}(x)  \quad \text{in $\Omega$}.
\end{equation}
\end{corollary}
Postponing a precise definition until Section \ref{sec:spectral}, let $A^{\ep}$ and $A^0$ be the self-adjoint operators corresponding to problems \eqref{resolventpde} and \eqref{eq:elastprob2}-\eqref{eq:elastprob3} respectively. We shall now state the second main result of this paper.
\begin{theorem}
\label{spectralcompactness}
The spectrum of $A^\ep$, $\sigma(A^\ep)$, converges in the sense of Hausdorff to the spectrum of $A^0$, $\sigma(A^0)$. That is 
\begin{enumerate}
\item For every $\lambda^0 \in \sigma(A^0)$ there exists $\lambda^\ep \in \sigma(A^\ep)$ such that $\lambda^\ep \rightarrow \lambda^0$.
\item If there exists $\lambda^\ep \in \sigma(A^\ep)$ such that $\lambda^\ep \rightarrow \lambda^0$, then $\lambda^0 \in \sigma(A^0)$.
\end{enumerate} 
\end{theorem}
Theorem \ref{spectralcompactness}, whose proof is given in Section \ref{sec:spectral}, tells us that if we wish to study the limit behaviour of the eigenvalues of $A^\ep$ as $\ep \rightarrow 0$ then it is sufficient to study the spectrum of the limit problem $\sigma(A^0)$. To this end consider $\sigma{(A^0)}$ and let $(\lambda^0,u^0)$ be an eigenvalue-eigenfunction pair of $A^0$. Then $u^0(x,y) = u(x) + v(x,y)$ satisfies
\begin{gather}
\label{eq:elastprob4} 
-\nabla \cdot \left( C^{\mathrm{hom}} \nabla u(x) \right)  = \lambda^0 u(x) + \lambda^0 \mv{v}(x)  \quad \text{in $\Omega$}, \\ \label{eq:elastprob5}
\begin{aligned}
 - \Delta_y v(x,y)   & = \lambda^0 v(x,y) + \nabla_y p(x,y) \quad \text{in $Q_2$} \\
 \nabla_y \cdot v(x,y) & = 0 \quad \text{in $Q_2$}   \\
 v(x,y) & = 0 \quad \text{on $\partial{Q_2}$}.
\end{aligned}
\end{gather}
Note that the equations \eqref{eq:elastprob4}-\eqref{eq:elastprob5} are not coupled: their uncoupled nature is due to the fact that $\lambda^0 u(x)$ originally present on the right hand side of \eqref{eq:elastprob5} can be absorbed by $p(x,y)$, i.e. $u(x) = \nabla_y q(x,y)$ for $q(x,y) = y \cdot u(x,y)$ for $y \in Q_2$. The main consequence of this uncoupling is that the spectrum of $A^0$ is simply the union of the spectra corresponding to problem \eqref{eq:elastprob4} and \eqref{eq:elastprob5} respectively, i.e.
\begin{corollary}
The spectrum of the homogenised limit operator $A^0$, $\sigma(A^0)$, has the following representation: 
$$
\sigma{(A^0)} = \{ \lambda_n \ \vert \  n \in \mathbb{N}  \} \cup \{ \mu_m \ \vert \ m \ \in \mathbb{N} \},
$$
where $\lambda_n$ satisfy, for some non-trivial $u_n \in [H^1_0(\Omega)]^d$,
$$
- \nabla \cdot \left( C^{\mathrm{hom}} \nabla u_n(x) \right)  = \lambda_n u_n(x)  \quad \text{in $\Omega$},
$$
and $\mu_m$ satisfy, for some non-trivial $v_m \in [H^1_0 (Q_2)]^d$, $p_m \in H^1 (Q_2)$,
\begin{align*}
 - \Delta_y v_m(y)    = \mu_m v_m(y) + \nabla_y p_m(y) \quad  & \text{in $Q_2$}, \\
 \nabla_y \cdot v_m(y)   = 0 \qquad\qquad\qquad  & \text{in $Q_2$}.   
\end{align*}
\end{corollary}

\begin{remark}
In the case of $\Omega = \mathbb{R}^d$ the operator $Lu : = - \nabla \cdot \left( C^{\mathrm{hom}} \nabla u(x) \right)$ has absolutely continuous spectrum coinciding with the positive real line. The spectrum of $A^0$ is still the union of the spectra for the operators defined by \eqref{eq:elastprob4} and \eqref{eq:elastprob5}, which implies that the spectrum of $A^0$ contains no gaps; $\sigma{(A^0)} = [0, \infty)$ and contains eigenvalues of infinite multiplicity $\mu_n$ corresponding to the eigenvalues of the Stokes problem on the inclusion $Q_2$. Furthermore, the spectral  convergence result, Theorem \ref{spectralcompactness}, can be shown to hold in this case too.
\end{remark}
\section{On two-scale convergence and its modifications for partial degeneracies}
\label{background}
Let us review the concept of two-scale convergence and some of its properties that are useful in the homogenisation of second order PDEs. For a full account of two-scale convergence and its application to homogenisation theory see e.g. \cite{nguetseng,allaire,zhikov1,zhikov2}. 
\subsection{Two-scale convergence and strong two-scale resolvent convergence}
\begin{definition} 
\label{dfn:wtwoscale}
\textbf{(Weak two-scale convergence.)} Let $u^{\ep}$ be a bounded sequence in $L^2(\Omega)$. We say $u^{\ep}$ (weakly) two-scale converges to $u^0 \in L^2(\Omega \times Q)$, denoted by $u^{\ep} \twoscale u^0$, if for all $\phi \in C^{\infty}_0(\Omega ; C^\infty_{\#}(Q))$, 
$$
\int_{\Omega} u^{\ep}(x)\phi\left(x, \tfrac{x}{\ep}\right) \mathrm{d}x \longrightarrow \int_{\Omega}\int_{Q} u^0(x,y)\phi(x , y) \ \mathrm{d}x\mathrm{d}y
$$
as $\ep \rightarrow 0$.
\end{definition}
The following properties of two-scale convergence are of particular importance to us:
\begin{lemma}[\textbf{Properties of (weak) two-scale convergence}] 
\label{twoscaleprop}
$ $
\begin{enumerate}[(i)]
\item If $u^{\ep}$ is bounded in $L^2(\Omega)$ then there exists $u^0 \in L^2(\Omega \times Q)$ and a subsequence $u^{\ep '}$ such that $u^{\ep '} \twoscale u^0$.
\item If $u^{\ep} \twoscale u^0$ then $u^{\ep}$ converges to $\int_{Q} u^0 \ \mathrm{d}y$ weakly in $L^2(\Omega)$.
\item If $u^{\ep} \twoscale u^0$ and $a(y) \in L^{\infty}(Q)$ then $a(x / \ep)u^{\ep}(x) \twoscale a(y)u^0(x,y)$.
\end{enumerate}
\end{lemma}
The following lemma is of particular importance to high contrast homogenisation:
\begin{lemma}\label{lem:pdhom1}
Let $u^\ep \in H^1(\Omega)$ such that 
\begin{align*}
\norm{u^\ep}{L^2(\Omega)} \le C, & & \ep \norm{\nabla u^\ep}{L^2(\Omega)} \le C,
\end{align*}
for some constant $C$ independent of $\ep$. Then, there exist $u^0(x,y)\in L^2(\Omega ; H^1_{\#}(Q))$ such that, up
to extracting a subsequence in $\varepsilon$ which we do not
relabel,
\begin{eqnarray}
u^\varepsilon&\stackrel{2}\rightharpoonup& u^0(x,y)         \label{eq:pdhom6} \\
\varepsilon \nabla u^\varepsilon  &\stackrel{2}\rightharpoonup& \nabla_y u^0(x,y).           \label{eq:pdhom7} 
\end{eqnarray}
\end{lemma}

In this paper we shall consider a sequence of non-negative self-adjoint operators $A^\ep$ and wish to study the limit behaviour of their resolvents, i.e. we wish to study the `resolvent' problem: for fixed $\alpha > 0$,
$$
A^\ep u + \alpha u = f,
$$
as $\ep \rightarrow 0$. In homogenisation theory we often find that $u^\ep : = (A^\ep + \alpha)^{-1} f \in L^2(\Omega)$ two-scale converges to some $u^0(x,y) \in L^2(\Omega \times Q)$ where 
$$
A^0 u^0 + \alpha u^0 = f,
$$
for a self-adjoint operator $A^0$. The notion of strong resolvent convergence is not applicable here since the limiting operator $(A^0 + \alpha)^{-1}$ is defined on a subspace of $L^2(\Omega \times Q)$ while $(A^\ep + \alpha)^{-1}$ is defined on $L^2(\Omega)$. We instead use the notion of strong two-scale resolvent convergence which we shall now recap, see \cite{zhikov1,zhikov2} for more details.
\begin{definition}\textbf{(Strong two-scale convergence.)}
\label{dfn:strongtwoscale}
A sequence $u^\ep \in L^2(\Omega)$ is said to strongly two-scale converge to $u \in L^2(\Omega \times Q)$, denoted $u^\ep \strongtwoscale u$, if $u^\ep \twoscale u$ and
$$
\int_{\Omega} u^\ep(x)v^\ep(x) \ \mathrm{d}x \rightarrow \int_{\Omega}\int_{Q} u(x,y)v(x,y) \ \mathrm{d}x\mathrm{d}y, \ \ \text{ for  all } v^\ep \twoscale v.
$$
Equivalently: $u^\ep \strongtwoscale u$ if, and only if,
\begin{align*}
u^\ep \twoscale u, & & \& & & \lim_{\ep \rightarrow 0}\int_{\Omega} \left( u^\ep \right)^2 (x) \ \mathrm{d}x = \int_{\Omega}\int_Q u^2(x,y) \ \mathrm{d}x\mathrm{d}y.
\end{align*}
\end{definition}
\begin{definition}
\label{app:twoscaleresolventcon}
Let $A^\ep$, $A^0$ be non-negative self-adjoint operators on $L^2(\Omega)$ and a closed linear subspace $H$ of $L^2(\Omega \times Q)$ respectively. Then we say $A^\ep$ strongly two-scale resolvent converges to $A^0$, denoted $A^\ep \strongtwoscale A^0$ if for some $\alpha > 0$ 
\begin{align*}
(A^\ep + \alpha)^{-1} f^\ep \strongtwoscale (A^0 + \alpha)^{-1} Pf \quad \text{whenever} \quad f^\ep \strongtwoscale f \in L^2(\Omega \times Q),
\end{align*}
where $P: L^2(\Omega \times Q) \rightarrow H$ is the orthogonal projection onto $H$.
\end{definition}
\begin{remark}
Since the resolvent set is an open subset of $\mathbb{C}$ and the resolvent is an analytic operator-valued function on $\mathbb{C}$ it is sufficient to test, as in the case of strong resolvent convergence, strong two-scale resolvent convergence for a single $\alpha>0$ in the resolvent set, say $\alpha = 1$.
\end{remark}
If $A^\ep$ strongly two-scale resolvent converges to $A^0$ then the limit spectrum $\sigma(A^0)$ is always contained in the limiting spectrum $\lim_{\ep \rightarrow 0}\sigma(A^\ep)$. That is
\begin{lemma}
\label{prop:app2s}
Let $A^\ep$, $A^0$ be non-negative self-adjoint operators on $L^2(\Omega)$ and the subspace $H \subset L^2(\Omega \times Q)$ respectively, let $A^\ep \strongtwoscale A^0$. Then for all $\lambda^0 \in \sigma(A^0)$ there exists $\lambda^\ep \in \sigma(A^\ep)$ such that $\lambda^\ep \rightarrow \lambda^0$ as $\ep \rightarrow 0$.
\end{lemma}
In general, it is not the case that the reverse inclusion holds. However if true, i.e. if $\lambda^\ep \in \sigma(A^\ep)$ such that $\lambda^\ep \rightarrow \lambda^0$ as $\ep \rightarrow 0$ implies $\lambda^0 \in \spec{A^0}$,  the proof is more difficult and problem specific. Such proofs usually require establishing, by separate means, a version of `two-scale spectral compactness', see Section \ref{sec:spectral}.

\subsection{Modification of two-scale convergence for partial degeneracies}
\label{subsec:degencell}
An important auxiliary problem in the homogenisation of problem \eqref{resolventpde} is the so called ``degenerate cell problem": find $v \in [H^1_{\#}(Q)]^d$ such that
\begin{align}
\label{auxiliaryprob}
 - \nabla_y \cdot \Big( C^{(1)}(y) \nabla_y v \Big) = F,
\end{align}
where $F \in H^{-1}_{\#}(Q)$ is given. $H^{-1}_{\#}(Q)$ is taken to be the dual space of $[H^1_{\#}(Q)]^d$ and we denote, for $F\in H^{-1}_{\#}(Q)$, by $\mv{F,w}$ the duality action of $F$ on $w \in [H^1_{\#}(Q)]^d$. The weak formulation of \eqref{auxiliaryprob} is: find $v\in \left[H_\#^1(Q)\right]^d$ such that
\begin{equation}
\int_Q\, C^{(1)}(y)\nabla_y v(y)\cdot \nabla_y w(y)\,dy\,= \mv{F,w},
\ \ \ \forall \ w\in \left[H_\#^1(Q)\right]^d.
\label{eq:pdhom19}
\end{equation} 

Introducing the space
\begin{equation}
V  := \left\{ u \in [H^1_{\#}(Q)]^d : C^{(1)}(y) \nabla_y u = 0 \right\}, \label{spaceV2}
\end{equation}
we see that if $v$ solves \eqref{auxiliaryprob}, equivalently \eqref{eq:pdhom19}, then $F$ must necessarily satisfy $\mv{F,u} = 0$ for all $u \in V$. This turns out, and was first shown by I.V. Kamotski and V.P. Smyshlyaev in \cite{ivkvps}, to be a sufficient criterion for solvability of \eqref{auxiliaryprob} when making the following {\bf Key assumption on the degeneracy:}
{\it There exists a constant $C>0$ such that for all $v\in \left[H_\#^1(Q)\right]^d$
there exists $v_1\in V$ with
\begin{equation}
\left\|v\,-v_1\right\|_{\left[H^1_\#(Q)\right]^d}\,\leq\, C\,\left\|C^{(1)}(y)\nabla_yv\right\|_2.
\label{eq:pdhom14}
\end{equation} }
\vspace{.1in}
The condition \eqref{eq:pdhom14} can be equivalently re-written as
\begin{equation}
\left\|P_{V^\bot} v\right\|_{\left[H^1_\#(Q)\right]^d}\,\leq\, C\,\left\| C^{(1)}(y)\nabla_yv\right\|_2,
\label{eq:pdhom15}
\end{equation}
where $P_{V^\bot}$ is the orthogonal projector in $\left[H^1_{\#}(Q)\right]^d$ on $V^\bot$, the orthogonal complement to $V$. The assumption \eqref{eq:pdhom14} does not depend on the choice of an equivalent norm in $H^1_{\#}$. The assumption \eqref{eq:pdhom14} holds for most of the previously studied cases, and has to be checked by separate means, see Section \ref{sec:spaceV}.

\begin{lemma}\label{lem:pdhom4} \ \\
\noindent (i) Under the assumption \eqref{eq:pdhom14}, the problem \eqref{eq:pdhom19} is solvable in
$\left[H_\#^1(Q)\right]^d$ if and only if
\begin{equation}
\mv{F,w}=\,0, \ \ \ \forall\, w\in V.
\label{eq:pdhom20}
\end{equation}
When \eqref{eq:pdhom20} does hold, the problem \eqref{eq:pdhom19}
is uniquely solvable in $V^\bot$. \newline
(ii) For any solution $v$ and any $v_1\in V$, $v+v_1$ is also a solution. Conversely,
any two solutions can only differ by a $v_1\in V$.
\end{lemma}
For completeness we will present the proof to Lemma \ref{lem:pdhom4} which was first proved in a greater generality by I.V. Kamotski and V.P. Smyshlyaev, cf \cite{ivkvps}.
\begin{proof}
(i)
Let $v$ be a solution of \eqref{eq:pdhom19} and let $w\in V$. Then,
using the symmetry of $C^{(1)}$ and \eqref{spaceV2},
\begin{equation}
\left\langle F\,,\, w\right\rangle\,=\,
 \int_Q\, C^{(1)}(y)\nabla_y
v(y)\cdot \nabla_y w(y)\,dy\,=\, \int_Q\, \nabla_y v(y)\cdot
C^{(1)}(y)\nabla_y w(y)\,dy\,=\,0 \label{fwsymm}
\end{equation}
 yielding
\eqref{eq:pdhom20}. Conversely let \eqref{eq:pdhom20} hold, and
seek $v\in \left[H_\#^1(Q)\right]^d$ solving
 \eqref{eq:pdhom19}.
 By \eqref{fwsymm}, the identity \eqref{eq:pdhom19} is automatically held for
all $w$ in $V$, therefore it is sufficient to verify it for all $w\in V^\bot$.
Seek $v$ also in $V^\bot$. Show that then, in the Hilbert space $H:=V^\bot$ with
the inherited $\left[H_\#^1(Q)\right]^d$ norm $\|\cdot\|_H$,
the problem   \eqref{eq:pdhom19} satisfies the conditions of the Lax-Milgram lemma, see for example \cite{evans}. Namely, first the bilinear form
\[
B[v,w]\,:=\,\int_Q\, C^{(1)}(y)\nabla_y v(y)\cdot \nabla_y w(y)\,dy
\]
is shown to be bounded in $H$, i.e. with some $C>0$,
\[
\biggl\vert\, B[\,v\,,\,w\,] \,\biggr\vert\,\leq\, C\,
\|v\|_H\,\|w\|_H, \ \ \ \forall v,\, w\,\in\,H.
\]
This follows from \eqref{elast:degentensor} where $C^{(2)} \in [L^\infty(Q)]^{d \times d \times d \times d}$. We will now show that the form $B$ is coercive, i.e. for some $\nu>0$,
\[
B[v,v]\,\geq \, \nu\, \|v\|_{H}^2, \ \ \ \forall v \in V^\bot.
\]
We have
\[
B[v,v]:=\int_Q\, C^{(1)}(y)\nabla_y v(y)\cdot \nabla_y v(y)\,dy=
\left\|\left(C^{(1)}(y)\right)^{1/2}\nabla_yv\right\|_2^2\,\geq\,
\]
\[
C\,\left\|\,C^{(1)}(y)\nabla_yv\,\right\|_2^2\,\,\geq\,\,
\nu\, \|v\|_H^2.
\]
In the last two inequalities we have used, the boundedness of $\left(C^{(1)}\right)^{1/2}$ and \eqref{eq:pdhom15}. 

Therefore, by the Lax-Milgram lemma, there exists a unique solution to the problem
\[
v\in V^\bot: \ \ B[v,w]\,= \mv{F,w}, \ \ \ \forall w\,\in\,V^\bot,
\]
and hence  to \eqref{eq:pdhom19}.

(ii) If $v$ solves \eqref{eq:pdhom19} and $v_1\in V$ then
$C^{(1)}(y)\nabla_yv_1(y)=0$ and hence $v+v_1$ also solves
\eqref{eq:pdhom19}. Assuming further $v^{(1)}$ and $v^{(2)}$ both solve \eqref{eq:pdhom19},
$v : =v^{(1)}-v^{(2)}$ solves \eqref{eq:pdhom19} with $F=0$, and then set
$w=v$. As a result,
\[
0\,=\,\int_Q\, C^{(1)}(y)\nabla_y v(y)\cdot \nabla_y v(y)\,dy\,=\,
\left\|\left(C^{(1)}(y)\right)^{1/2}\nabla_yv\right\|_2^2,
\]
implying $\left(C^{(1)}\right)^{1/2}\nabla_yv\,=\,0$ and hence
$C^{(1)}\nabla_yv\,=\,0$, i.e. $v\in V$.
\end{proof}

\section{Homogenisation}
\label{sec:hom}
\subsection{Proof of Theorem \ref{thm:elastprob2}}
The weak form of \eqref{resolventpde} is stated as follows: Find $u^\ep \in [H^1_0 (\Omega)]^d$ such that
\begin{multline}
\label{eq:elasthom1}
\int_{\Omega} C^{(1)}\left(\tfrac{x}{\ep}\right) e(u^\ep) \cdot e(\phi) + \ep^2 C^{(0)}\left(\tfrac{x}{\ep}\right) e(u^\ep) \cdot  e(\phi) + \alpha u^\ep \cdot \phi = \int_{\Omega} f^{\ep} \cdot \phi \\ \forall \phi \in [H^1_0(\Omega)]^d.
\end{multline}
For any fixed $\ep > 0,$ the associated quadratic form 
\begin{equation}
\label{abc1}
\mathcal{A}^\ep(u,v) : = \int_{\Omega} C^{(1)}\left(\tfrac{x}{\ep}\right) e(u) \cdot e(v) + \ep^2 C^{(0)}\left(\tfrac{x}{\ep}\right) e(u) \cdot  e(v) + \alpha u \cdot v,
\end{equation}
 is coercive, i.e. there exists $C>0$ such that
$$
\mathcal{A}^\ep(u,v) \ge C \norm{u}{H^1_0(\Omega)}^2, \quad \forall u \in H^1_0(\Omega).
$$
This is established by noting $C^{(1)} + C^{(0)}$ is positive definite, see \eqref{eq:elasthom2}, and an application of the standard Korn's inequality, see e.g \cite{CPS}. By the Lax-Milgram lemma, c.f. \cite{evans}, coercivity ensures the existence and uniqueness of $u^\ep \in [H^1_0(\Omega)]^d$ for fixed $\alpha \ge0$. We wish to study how $u^\ep$ behaves as $\ep$ tends to zero. Substituting $\phi = u^{\ep}$ into \eqref{eq:elasthom1} and using \eqref{eq:elasthom2} we see that there exists a constant $C$ independent of $\ep$ such that
\begin{align}
\label{elast:aprioribounds1}
\norm{u^{\ep}}{L^2(\Omega)} & \le C\norm{f^\ep}{L^2(\Omega)}, \\ \norm{\ep \nabla u^{\ep}}{L^2(\Omega)} & \le C\norm{f^\ep}{L^2(\Omega)}, \label{elast:aprioribounds2} \\ \norm{(C^{(1)}(x / \ep))^{1/2}\nabla u^{\ep}}{L^2(\Omega)} & \le C\norm{f^\ep}{L^2(\Omega)}. \label{elast:aprioribounds3}
\end{align}
Note that for $\alpha = 0$ we require the validity of the following Poincar\'{e} type inequality: There exists a constant $C>0$ independent of $\ep$ such that for all $u \in H^1_0(\Omega)$
$$
\norm{u}{L^2(\Omega)}^2 \le C \left( \norm{C^{(1)}(x/\ep) \nabla u}{L^2(\Omega^\ep_1)}^2 + \ep^2 \norm{C^{(0)}(x/\ep) \nabla u}{L^2(\Omega^\ep_2)}^2  \right),
$$
which indeed holds, the proof of which is given in Section \ref{subsec:spec}, Lemma \ref{lem:epti}. The uniform bounds \eqref{elast:aprioribounds1}-\eqref{elast:aprioribounds3}, and the two-scale compactness result, see Lemma \ref{twoscaleprop} (i), imply that, for uniformly bounded $f^\ep$, these sequences have two-scale convergent subsequences and the behaviour of their two-scale limits is described in the following theorem:

\begin{theorem}
\label{lem:elasthom1}
There exists $u(x)\in [H^1_0(\Omega)]^d$, $ v(x,y) \in L^2(\Omega ;[H^1_0(Q_2)]^d )$, $\nabla_y \cdot v(x,y)=0$, such that, up to a subsequence in $\ep$ (which we do not relabel),
\begin{flalign*}
u^{\ep} & \twoscale u(x) + v(x,y), \\
\ep \nabla u^{\ep} & \twoscale \nabla_y v(x,y), \\
( C^{(1)})^{1/2}(x / \ep) \nabla u^{\ep} & \twoscale (C^{(1)})^{1/2}(y)\left[ \nabla_x u(x) + \nabla_y u^1(x,y) \right] ,
\end{flalign*}
where $u^1(x,y) \in [L^2(\Omega ; H^1_{\#}(Q))]^d$ is a solution to
\begin{equation}
\label{eq:elasthom9}
-\nabla_y \cdot \left( C^{(1)}(y) \nabla_y u^1(x,y) \right) = \nabla_y \cdot \left( C^{(1)}(y) \nabla_x u(x) \right).
\end{equation}
\end{theorem}
\begin{proof} \\
\textbf{Step 1:} By \eqref{elast:aprioribounds1}-\eqref{elast:aprioribounds2} and Lemma \ref{lem:pdhom1}, up to a subsequence in $\ep$, which we do not relabel, $u^\ep$ and $\ep \nabla u^{\ep}$ two-scale converge to $u^0$ and $\nabla_y u^0$ respectively.

Let us show $u^0 \in [L^2(\Omega; V)]^d$, for $V$ given by \eqref{spaceV2}. Since $\ep \nabla u^\ep \twoscale \nabla_y u^0$, by Definition \ref{dfn:wtwoscale} and Lemma \ref{twoscaleprop} (iii), 
 \begin{multline*}
\int_{\Omega} \ep \Big(C^{(1)}\left(\tfrac{x}{\ep}\right)\Big)^{1/2} \nabla u^{\ep}(x)\cdot \phi\left(x,\tfrac{x}{\ep}\right) \ \mathrm{d}x \\ = \int_{\Omega} \ep \nabla u^{\ep}(x) \cdot \Big(C^{(1)}\left(\tfrac{x}{\ep}\right)\Big)^{1/2} \phi\left(x,\tfrac{x}{\ep}\right) \ \mathrm{d}x \\
\longrightarrow \int_{\Omega}\int_{Q} \nabla_y u^0(x,y) \cdot \Big(C^{(1)}\left( y \right)\Big)^{1/2} \phi(x,y)   \ \mathrm{d}y\mathrm{d}x \\ = \int_{\Omega}\int_{Q} \Big(C^{(1)}\left( y \right)\Big)^{1/2} \nabla_y u^0(x,y) \cdot \phi(x,y)   \ \mathrm{d}y\mathrm{d}x, \quad \forall \phi \in [C^\infty_0(\Omega ; C^\infty_\#(Q)]^d.
\end{multline*}
By \eqref{elast:aprioribounds3}, $(C^{(1)}(x/\ep))^{1/2} \nabla u^{\ep}(x)$ is bounded in $L^2(\Omega)$ so 
$$
\int_{\Omega} \ep \Big(C^{(1)}\left(\tfrac{x}{\ep}\right)\Big)^{1/2} \nabla u^{\ep}(x) \cdot \phi\left(x,\tfrac{x}{\ep}\right) \ \mathrm{d}x  \longrightarrow 0.
$$
Therefore
\begin{equation*}
\int_{\Omega}\int_{Q} (C^{(1)}(y))^{1/2} \nabla_y u^0(x,y) \cdot \phi(x,y)   \ \mathrm{d}y\mathrm{d}x = 0.
\end{equation*}
This implies, since the functions $\phi(x,y)$ are dense in $[L^2(\Omega \times Q)]^{d \times d}$, that 
\begin{equation}
\label{proofofu0inV}
(C^{(1)}(y))^{1/2} \nabla_y u^0(x,y) = 0 \quad \text{ a.e. } x \in \Omega.
\end{equation} 
Premultiplying \eqref{proofofu0inV} by $(C^{(1)}(y))^{1/2}$ shows $u^0(x,y) \in L^2(\Omega;V)$. For a.e. $x \in \Omega$, $u^0(x, \cdot) \in V$, and by \eqref{eq:elastV1} below, we see $u^0(x,y)= u(x) + v(x,y)$ for some $u \in [L^2(\Omega)]^d$, $v \in [L^2(\Omega ; H^1_0(Q_2))]^d$ with $\nabla_y \cdot v =0$. 

Inequality \eqref{elast:aprioribounds3} and Lemma \ref{twoscaleprop} (i) imply that there exists $\xi^0 \in [L^2(\Omega \times Q)]^{d \times d}$ such that, up to a subsequence in $\ep$ which we do not relabel, $(C^{(1)}(x/\ep))^{1/2} \nabla u^{\ep}(x) \twoscale \xi^0(x,y)$. Introducing the space
\begin{equation}
\label{spaceW}
W : = \left\{ \Psi \in [L^2(Q) ]^{d \times d} : \text{$\nabla_y \cdot \left(\left( C^{(1)}(y) \right)^{1/2}\Psi(y)\right)=0$ in $H^{-1}_{\#}(Q)$}\right\} ,
\end{equation}
we will show that $\xi^0 \in L^2(\Omega ; W)$. Take in \eqref{eq:elasthom1}
$\phi(x)=\phi^\varepsilon(x)=\,
\varepsilon\,\Phi\left(x,\frac{x}{\varepsilon}\right)$ for
any $\Phi(x,y)\in \left[C^\infty_0\left(\Omega;
C_\#^\infty(Q)\right)\right]^{d}$. Passing then to the limit
in \eqref{eq:elasthom1} we notice, via \eqref{elast:aprioribounds1}-\eqref{elast:aprioribounds3}, that the limit of each term but the first one
on the left hand-side of \eqref{eq:elasthom1} is zero, and therefore
\begin{multline*}
\lim_{\varepsilon \to 0} \int_\Omega
C^{(1)}\left(\frac{x}{\varepsilon}\right) \nabla
u^\varepsilon(x)\,\cdot\,\varepsilon\nabla \Phi\left(x,
\frac{x}{\varepsilon}\right) \,dx\, \\ =\, \int_\Omega\int_Q
\left(C^{(1)}(y)\right)^{1/2}
\xi^0(x,y)\,\cdot\,\nabla_y\Phi(x,y)\,dy\,dx,=\,0.
\end{multline*}
The density of $\Phi(x,y)$ implies then that, for a.e. $x$,
$$
\nabla_y \cdot \Big( \left(C^{(1)}(y)\right)^{1/2}\xi^0(x,y) \Big) = 0 \qquad \text{in $H^{-1}_{\#}(Q)$.}
$$
This yields $\xi^0(x,y)\in L^2\left(\Omega;\,W\right)$, see \eqref{spaceW}. Hence, we have shown
\begin{flalign*}
u^{\ep} & \twoscale u(x) + v(x,y), \\
\ep \nabla u^{\ep} & \twoscale \nabla_y v(x,y), \\
( C^{(1)})^{1/2}(x / \ep) \nabla u^{\ep} & \twoscale \xi^0(x,y) ,
\end{flalign*}
for some $u \in [L^2(\Omega)]^d$, $v(x,y) \in L^2(\Omega ; [H^1_0(Q_2)]^d)$, $\nabla_y \cdot v=0$ and $\xi^0(x,y) \in L^2(\Omega ; W)$.

\textbf{Step 2:} Let us now show that $u(x) \in [H^1_0(\Omega)]^d$. For $\Psi \in C^{\infty}(\Omega ; W)$ we see via integration by parts 
$$
\int_{\Omega} \Big(C^{(1)}\left(\tfrac{x}{\ep}\right)\Big)^{1/2} \nabla u^{\ep}(x) \cdot \Psi\left(x, \tfrac{x}{\ep}\right) \ \mathrm{d}x = - \int_{\Omega} u^{\ep} \cdot \left( \nabla_x \cdot \Big(C^{(1)}\left(\tfrac{x}{\ep}\right)\Big)^{1/2} \Psi\left(x, \tfrac{x}{\ep}\right)  \right)
$$
and passing to the two-scale limit indicates $u^0(x,y)$ and $\xi^0(x,y)$ are related by the following expression
\begin{equation}
\label{fluxproj}
\begin{split}
\int_{\Omega}\int_{Q} \xi^0(x,y) \cdot \Psi(x,y) \ \mathrm{d}y\mathrm{d}x = - \int_{\Omega}\int_{Q} u^0(x,y) \cdot \nabla_x \cdot\left( (C^{(1)}(y))^{1/2}\Psi(x,y) \right) \ \mathrm{d}y\mathrm{d}x  \\ \forall \Psi \in C^{\infty}(\Omega;W).
\end{split}
\end{equation} 
Now, for fixed $\varphi \in [C^\infty(\Omega)]^{d \times d}$ we take in \eqref{fluxproj} test functions $\Psi_\varphi$ given by Lemma \ref{prop:xreg} at the end of this section. Then, \eqref{fluxproj} states, via \eqref{xreq1},
\begin{equation}
\label{elast:weakderivative1}
\ell(\varphi) = - \int_{\Omega} u(x) \cdot \nabla_x \cdot \varphi \ \mathrm{d}x,
\end{equation}
where
$$
\ell(\varphi) : = \int_{\Omega}\int_{Q} \xi_0(x,y) \cdot \Psi_\varphi(x,y) \ \mathrm{d}y\mathrm{d}x, 
$$
defines an $L^2$ bounded linear form on $\varphi \in [C^\infty(\Omega)]^{d \times d}$, see \eqref{xreq2}. Therefore, by the Riesz representation theorem, \eqref{elast:weakderivative1} implies
$$
\ell(\phi) = \int_\Omega v \cdot \varphi \ \mathrm{d}x,
$$
for some $v \in [L^2(\Omega)]^{d \times d}$ and hence, via \eqref{elast:weakderivative1}, $u \in [H^1_0(\Omega)]^d$. 

\textbf{Step 3:} It remains to show $\xi^0(x,y) = (C^{(1)})^{1/2}(y)[ \nabla_x u(x) + \nabla_y u^1(x,y)]$ for some $u^1$ given by \eqref{eq:elasthom9}. For a.e. $x \in \Omega$, let $u^1(x, \cdot) \in [H^1_{\#}(Q)]^d$ be a solution of \eqref{eq:elasthom9}. Note that, if the key assumption \eqref{eq:pdhom14} holds, such a solution exists by Lemma \ref{lem:pdhom4}  since, for $F: = \nabla_y \cdot \left( C^{(1)}(y) \nabla_x u(x) \right)$,
\begin{multline*}
\mv{F,v} = - \int_Q C^{(1)}(y) \nabla_x u(x) \cdot \nabla_y v \ \mathrm{d}y = - \int_Q \nabla_x u(x) \cdot C^{(1)}(y) \nabla_y v \ \mathrm{d}y = 0 \\ \forall v \in V.
\end{multline*}
The key assumption \eqref{eq:pdhom14} does indeed hold and the proof of this fact is the subject of Section \ref{sec:spaceV}. Setting $\xi(x,y) : = (C^{(1)})^{1/2}[ \nabla_x u(x) + \nabla_y u^1(x,y)]$, we note $\xi$ is well-defined, since $u^1(x,\cdot)$ is unique up to a function in $V$. Furthermore, $\xi(x,y) \in [L^2(\Omega ; W)]^d$, from \eqref{eq:elasthom9}, with
\begin{equation}
\label{fluxproj2}
\begin{split}
\int_{\Omega}\int_{Q} \xi(x,y) \cdot \Psi(x,y) \ \mathrm{d}y\mathrm{d}x = - \int_{\Omega}\int_{Q} u(x) \cdot \nabla_x \cdot \left( (C^{(1)}(y))^{1/2}\Psi(x,y) \right) \ \mathrm{d}y\mathrm{d}x \quad  \\ \forall \Psi \in C^{\infty}(\Omega;W).
\end{split}
\end{equation} 
The latter directly follows via integration by parts and \eqref{spaceW}. Now let us show that $\xi_0(x,y) = \xi(x,y)$ for a.e. $x\in \Omega$: since both $\xi_0(x,\cdot)$ and $\xi(x,\cdot)$ belong to $W$, it is sufficient to show   $\xi_0(x,\cdot) - \xi(x,\cdot) \perp W$. To this end, \eqref{fluxproj} and \eqref{fluxproj2} imply that 
\begin{multline}
\label{eq:elasthom8}
\int_{\Omega}\int_{Q} \left( \xi_0(x,y) - \xi(x,y) \right)\cdot \Psi(x,y) \ \mathrm{d}y\mathrm{d}x = \\ - \int_{\Omega}\int_{Q} v(x,y) \cdot \nabla_x \cdot \left( (C^{(1)}(y))^{1/2}\Psi(x,y) \right) \mathrm{d}y\mathrm{d}x \quad  \forall \Psi \in C^\infty(\Omega ;W). \end{multline}
Now the result follows if the right hand side of \eqref{eq:elasthom8} is zero. Since $C^{\infty}_{0}(\Omega ; V)$ is dense in $L^2(\Omega ; V)$ it is sufficient to show that for a fixed $\phi \in [C^{\infty}_{0}(\Omega;H^1_0(Q_2) \cap V)]^d$ 
\begin{equation}
\label{eq:elasthom12}
\int_{\Omega} \int_{Q_2} (C^{(1)}(y))^{1/2} \nabla_x \phi(x,y) \cdot \Psi\ \mathrm{d}y\mathrm{d}x =0 \quad \forall \Psi \in C^\infty(\Omega; W)
\end{equation}
The latter can be seen by considering a function $N \in [H^1_{\#}(Q)]^d$ such that $N(y)=y$ for $y \in Q_2$. An example of such a function would be an appropriate periodic extension of $f(y)=\chi_2(y) y$, which is possible since $\overline{Q_2} \subset (0,1)^d$. Now for given $\phi$ define an auxiliary function $\Phi(x,y) : = \nabla_x \phi(x,y) \cdot N(y)$. For a.e. $x \in \Omega$,   $\Phi(x,\cdot) \in [H^1_{\#}(Q)]^d$, and for any $\Psi \in W$
\begin{align*}
0 & = \mv{- \nabla_y \cdot \Big((C^{(1)}(y))^{1/2} \Psi\Big),\Phi(x,y)} = \int_{Q} \Psi(y) \cdot (C^{(1)}(y))^{1/2} \nabla_y \Phi(x,y) \ \mathrm{d}y \\
& = \int_{Q} \Psi(y) \cdot (C^{(1)}(y))^{1/2} \nabla_y (\nabla_x \phi(x,y) \cdot N(y)) \ \mathrm{d}y \\ 
& = \int_{Q_2} \Psi(y) \cdot (C^{(1)}(y))^{1/2} \nabla_y (\nabla_x \phi(x,y) \cdot y) \ \mathrm{d}y \\ 
& = \int_{Q_2} \Psi(y) \cdot (C^{(1)}(y))^{1/2} \nabla_x \phi(x,y) \ \mathrm{d}y,
\end{align*}
where the last equality is due to the fact that, for $\phi(x,\cdot) \in V$, in components,
\begin{flalign*}
\Big((C^{(1)}(y))^{1/2} \nabla_y (\nabla_x \phi(x,y) \cdot y)\Big)_{ij} & = (C^{(1)}(y))^{1/2}_{ijpq}( \phi_{p,x_s}y_s )_{,y_q} \\
&  = (C^{(1)}(y))^{1/2}_{ijpq}( \phi_{p,x_s y_q}y_s + \phi_{p,x_s} \delta_{sq} ) \\
& = \underbrace{(C^{(1)}(y))^{1/2}_{ijpq}( \phi_{p,x_s y_q}y_s )}_{= 0} + (C^{(1)}(y))^{1/2}_{ijpq}( \phi_{p,x_q} ) \\
& = \Big((C^{(1)}(y))^{1/2} \nabla_x \phi(x,y)\Big)_{ij}.
\end{flalign*}
Here, we have used the notation $,z$ to denote differentiation with respect to the variable $z$, for example $\phi_{p,x_q} : = \pd{\phi_p}{x_q}$.
\end{proof}
Using Theorem \ref{lem:elasthom1} we are now able to pass to the limit as $\ep \rightarrow 0$ in \eqref{eq:elasthom1} and prove Theorem \ref{thm:elastprob2}.
\\

\noindent \begin{proof} \textbf{of Theorem \ref{thm:elastprob2}.}

\textbf{Step 1:} For fixed $\phi(x,y) \in C^{\infty}_{0}(\Omega ; V)$, employing the test function $\phi^0(x) := \phi(x,\tfrac{x}{\ep})$ in \eqref{eq:elasthom1} gives
\begin{multline}
\label{eq:elasthom10}
\int_{\Omega} C^{(1)}\left(\tfrac{x}{\ep}\right)  \nabla u^\ep \cdot \nabla_x \phi\left( x, \tfrac{x}{\ep}\right) +  \ep C^{(0)}\left(\tfrac{x}{\ep}\right)\nabla u^\ep \cdot \left[ \ep \nabla_x \phi\left( x, \tfrac{x}{\ep}\right) + \nabla_y \phi\left( x, \tfrac{x}{\ep}\right) \right]  + \\ + \int_{\Omega} \alpha u^\ep \cdot \phi\left( x, \tfrac{x}{\ep}\right) = \int_{\Omega} f^{\ep} \cdot \phi\left( x, \tfrac{x}{\ep}\right).
\end{multline}
Passing in \eqref{eq:elasthom10} to the limit $\ep \rightarrow 0$, using Theorem \ref{lem:elasthom1}, gives
\begin{multline}
\label{eq:elasthom11}
\int_{\Omega}\int_{Q} C^{(1)}(y) \left[ \nabla_x u(x) + \nabla_y u^1(x,y) \right] \cdot \nabla_x \phi(x,y) \ \mathrm{d}y\mathrm{d}x  + \\ + \int_{\Omega}\int_Q C^{(0)}(y)\nabla_y v(x,y) \cdot \nabla_y \phi(x,y) + \alpha \left( u(x) + v(x,y) \right)\cdot \phi(x,y) \ \mathrm{d}y\mathrm{d}x  \\ = \int_{\Omega}\int_Q f(x,y) \phi(x,y) \ \mathrm{d}y\mathrm{d}x , \quad \forall \phi(x,y) \in [C^{\infty}_{0}(\Omega ; V)]^d.
\end{multline}

\textbf{Step 2:} Choosing in \eqref{eq:elasthom11} $\phi(x,y) \equiv \varphi(x)$, $\varphi \in C^{\infty}_0(\Omega)$, gives
\begin{multline*}
\int_{\Omega}\int_{Q} C^{(1)}(y)( \nabla_x u(x) + \nabla_y u^1(x,y) ) \cdot \nabla_x \varphi(x) + \alpha \left(u(x) + v(x,y)\right)\varphi(x) \ \mathrm{d}y\mathrm{d}x \\ = \int_{\Omega}\int_{Q} f(x,y) \varphi(x) \ \mathrm{d}y\mathrm{d}x, 
\end{multline*}
which is the variational form for
\begin{equation}
\label{1homlimit} 
- \nabla \cdot \left( \left< C^{(1)} ( \nabla_x u + \nabla_y u^1) \right> (x) \right)   + \alpha \left( u(x) + <v>(x) \right)  = <f>(x)  \quad \text{in $\Omega$}.
\end{equation}
Setting $u^1_p = N^p_{rs}(y)\pd{u_r}{x_s}(x)$ and substituting into \eqref{eq:elasthom9} and \eqref{1homlimit} gives equations \eqref{eq:elastprob2} and \eqref{elastinserted1}-\eqref{eq:ehom0.2}.  Let us show that $C^{\text{hom}}$ is strictly positive. For any  symmetric $\eta \in \mathbb{R}^{d \times d}$ we can show, as in the case of the perforated domain (see e.g. \cite{zhikov3}), $C^{\text{hom}} \eta \cdot \eta$ has the following variational representation
$$
C^{\text{hom}} \eta \cdot \eta = \inf_{w \in [C^{\infty}_{\#}(Q)]^d} \int_{Q} C^{(1)}(y)\left( \eta + \nabla_y w  \right) \left(  \eta + \nabla_y w \right) \ \mathrm{d}y.
$$

By \eqref{elast:degentensor} it is easy to see that
\begin{flalign*}
C^{\text{hom}} \eta \cdot \eta  & \ge \inf_{w \in [C^{\infty}_{\#}(Q)]^d} \int_{Q_1} C^{(2)}(y)\left( \eta + \nabla_y w  \right) \left(  \eta + \nabla_y w \right) \ \mathrm{d}y \\ 
& = \hat{C} \eta \cdot \eta,
\end{flalign*}
where $\hat{C}$ is the homogenised tensor for the perforated linear elasticity problem which is well known to be strictly positive, see e.g. \cite{zhikov3}. Therefore $C^{\text{hom}}$ is strictly positive.

\textbf{Step 3:} Now choose in \eqref{eq:elasthom11} the test functions of the form $\phi_0(x,y) = \psi(x)\phi(y)$, $\psi \in C^{\infty}_0(\Omega)$, $\phi \in C^{\infty}_0(Q_2)$ with $\nabla \cdot \phi =0$. Due to \eqref{eq:elasthom12} and the fact $\xi_0(x,y) = \big(C^{(1)}(y)\big)^{1/2} \left[ \nabla_x u(x) + \nabla_y u^1(x,y) \right] \in [L^2(\Omega ; W)]^d$, we have
$$
\int_{\Omega}\int_Q (C^{(1)}(y)) \left[ \nabla_x u(x) + \nabla_y u^1(x,y) \right] \nabla_x \psi(x)\phi(y) \ \mathrm{d}y\mathrm{d}x = 0
$$
Furthermore
\begin{multline*}
\int_{\Omega}\int_{Q_2} C^{(0)}(y)\nabla_y v(x,y) \cdot \nabla_y\phi(y) \ \mathrm{d}y\mathrm{d}x = \int_{\Omega}\int_{Q_2} ( \delta_{ip}\delta_{jq}  + \delta_{iq}\delta_{jp} )v_{p,y_q}\phi_{i,y_j} \ \mathrm{d}y\mathrm{d}x \\
=  \int_{\Omega}\int_{Q_2} ( v_{p,y_q}\phi_{p,y_q}  + v_{p,y_q}\phi_{q,y_p} ) \ \mathrm{d}y\mathrm{d}x = \int_{\Omega}\int_{Q_2} ( v_{p,y_q}\phi_{p,y_q}  + v_{p,y_p}\phi_{q,y_q} ) \ \mathrm{d}y\mathrm{d}x \\ 
= \int_{\Omega}\int_{Q_2} \nabla_y v \cdot \nabla_y \phi  \ \mathrm{d}y\mathrm{d}x.
\end{multline*}
Therefore, 
\begin{multline*}
\int_{\Omega}\int_{Q_2} \nabla_y v(x,y) \cdot \nabla_y \phi(y)\psi(x) + \alpha \left( u(x) + v(x,y) \right)\cdot \psi(x)\phi(y) \ \mathrm{d}y\mathrm{d}x  \\ = \int_{\Omega}\int_{Q_2} f(x,y) \psi(x)\phi(y) \ \mathrm{d}y\mathrm{d}x,
\end{multline*}
which is the variational form for \eqref{eq:elastprob3}. 

\textbf{Step 4:} It remains to show that if $f^\ep \strongtwoscale f$ then $u^\ep \strongtwoscale u^0$. This proof is similar to the double porosity case, see e.g. \cite{zhikov1}, and we will present it here for completeness. Let $z^\ep \in [H^1_0(\Omega)]^d$ be the solution to
\begin{equation}
\label{eq:pdelaststrongtwoscale1}
- \nabla \cdot \left( C^\ep \nabla z^\ep \right) + \alpha z^\ep = u^\ep. 
\end{equation}
Then, by the above arguments, $z^\ep \twoscale z^0$ where $z^0(x,y) = z(x) + w(x,y)$ is a solution to
\begin{multline}
\label{eq:pdelaststrongtwoscale2}
\int_{\Omega}\int_{Q} C^{(1)}_{ijpq}(y) \left[ \delta_{pr}\delta_{qs} + N^p_{rs,q}(y) \right] z_{r,s}(x)  \phi_{i,x_j}(x,y) \ \mathrm{d}y\mathrm{d}x  + \\  + \int_{\Omega}\int_Q C^{(0)}(y)\nabla_y w(x,y) \cdot \nabla_y \phi(x,y) + \alpha \left( z(x) + w(x,y) \right)\cdot \phi(x,y) \ \mathrm{d}y\mathrm{d}x  \\ = \int_{\Omega}\int_Q u^0(x,y) \phi(x,y) \ \mathrm{d}y\mathrm{d}x , \quad \forall \phi(x,y) \in [C^{\infty}_{0}(\Omega ; V)]^d.
\end{multline}
Setting $\phi = z^0$ in \eqref{eq:elasthom11} and $\phi = u^0$ in \eqref{eq:pdelaststrongtwoscale2} shows
\begin{equation}
\label{eq:pdelaststrongtwoscale3}
\dblint{\Omega}{ \left\vert u^0(x,y) \right\vert^2} = \dblint{\Omega}{f(x,y) z^0(x,y)}.
\end{equation}
Similarly, the variational forms for \eqref{eq:pdelaststrongtwoscale1} and \eqref{resolventpde} show
\begin{equation}
\label{eq:pdelaststrongtwoscale4}
\int_\Omega \vert u^\ep(x) \vert^2 \ \mathrm{d}x = \int_\Omega f^\ep(x) \cdot z^\ep(x) \ \mathrm{d}x.
\end{equation}
Using the assumption $f^\ep \strongtwoscale f^0$, passing to the limit $\ep \rightarrow 0$ in \eqref{eq:pdelaststrongtwoscale4}, via \eqref{eq:pdelaststrongtwoscale3}, gives
\begin{multline*}
\lim_{\ep \rightarrow 0}\int_\Omega \vert u^\ep \vert^2 \ \mathrm{d}x =\lim_{\ep \rightarrow 0}\int_\Omega f^\ep(x) \cdot z^\ep(x) \ \mathrm{d}x = \dblint{\Omega}{f(x,y) \cdot z^0(x,y)} \\ = \dblint{\Omega}{ \vert u^0(x,y) \vert^2}.
\end{multline*}
Hence, by Definition \ref{dfn:strongtwoscale}, $u^\ep \strongtwoscale u^0$.

Finally, observe that the uniqueness to the solution of \eqref{eq:elastprob2}-\eqref{eq:elastprob3} follows in a standard way by setting in \eqref{eq:elasthom11} $\phi = u^0$ and $f = 0$.
\end{proof}

\begin{proof}\textbf{of Corollary \ref{cor:elastprob1}.}
The equation for the microscopic deformations $v(x,y)$ is given by
\begin{multline}
\label{eq:elasthom13}
\int_{\Omega}\int_{Q_2} \nabla_y v(x,y) \cdot \nabla_y\psi(y) \phi(x) + \alpha \Big( u(x) + v(x,y) \Big)\cdot \phi(x)\psi(y) \ \mathrm{d}y\mathrm{d}x  \\ = \int_{\Omega}\int_{Q_2} f(x,y) \phi(x)\psi(y) \ \mathrm{d}y\mathrm{d}x,
\end{multline}
for all $\phi \in C^{\infty}_0(\Omega)$, $\psi \in C^{\infty}_0(Q_2)$ with $\nabla \cdot \psi=0$. For an externally applied body force $f^{\ep}(x) = f(x, x / \ep)$, where $f(x,y) = f_0(x) + \nabla_y f_1 (x,y)$, such that $f^\ep \strongtwoscale f(x,y)$ as $\ep \rightarrow 0$ we find, by taking into account that for any constant vector field $c$ we can have the representation $c = \nabla_y ( c \cdot y)$ in $L^2(Q_2)$, that 
\begin{flalign*}
\int_{\Omega}\int_{Q_2} f(x,y) \cdot \phi(x)\psi(y) & \ \mathrm{d}y\mathrm{d}x  =
\int_{\Omega}\int_{Q_2} \left[ f_0(x) + \nabla_y f_1 (x,y) \right] \phi(x)\psi(y) \ \mathrm{d}y\mathrm{d}x \\
 & = \int_{\Omega}\int_{Q_2} \left[ \nabla_y \left( f_0(x) \cdot y \right)  + \nabla_y f_1 (x,y) \right] \phi(x)\psi(y) \ \mathrm{d}y\mathrm{d}x \\
& = - \int_{\Omega}\int_{Q_2} \left[ f_0(x) \cdot y +  f_1 (x,y) \right] \phi(x) \nabla_y \cdot \psi(y) \ \mathrm{d}y\mathrm{d}x = 0,
\end{flalign*}
for all $\psi \in C^{\infty}_0(Q_2)$ with $\nabla \cdot \psi =0$. Similarly, $\int_{\Omega}\int_{Q_2}  u(x) \cdot \phi(x)\psi(y) \ \mathrm{d}y\mathrm{d}x = 0.$  Therefore equation \eqref{eq:elasthom13} becomes 
\begin{equation*}
\int_{\Omega}\int_{Q_2} \nabla_y v(x,y) \cdot \nabla_y\psi(y) \phi(x) + \alpha v(x,y) \cdot \phi(x)\psi(y) \ \mathrm{d}y\mathrm{d}x =0.
\end{equation*}
This implies that for a.e. $x$ in $\Omega$, $v(x,\cdot)$ is a weak solution of the homogeneous Stokes problem which is well known to have only the trivial solution $v(x,y)=0$, see e.g. \cite{temam}. Furthermore, as $u^\ep \strongtwoscale u(x)$, $u^\ep \rightarrow u$ strongly in $L^2$. 
\end{proof}
Let us end the section with the proof of the Lemma used to show the regularity of the limit function $u(x)$.
\begin{lemma}
\label{prop:xreg}
For fixed $\varphi \in [C^\infty(\Omega)]^d$, there exists $\Psi(x,y) \in C^\infty(\Omega ; W)$, such that $\Psi \equiv 0$ in $Q_2$, and
\begin{equation}
\label{xreq1}
\nabla_x \cdot \varphi = \nabla_x \cdot \left( \int_{Q_1} \left( C^{(1)}(y) \right)^{1/2} \Psi(x,y) \ \mathrm{d}y \right)
\end{equation}
Furthermore, there exists a constant $C$ independent of $\varphi$ such that
\begin{equation}
\label{xreq2}
\left\vert \int_\Omega \int_Q \xi \cdot \Psi \right\vert \le C \left( \int_\Omega \left\vert \varphi \right\vert^2 \right)^{1/2}\left( \dblint{\Omega}{ \left\vert \xi \right\vert^2} \right)^{1/2}, \quad \ \forall \xi \in L^2(\Omega ; W).
\end{equation}
\end{lemma}
\begin{proof}
Let us introduce the tensor
$$
\hat{C}_{ijrs} : = \int_{Q_1} C^{(2)}_{ijpq}(y) \left( \delta_{pr}\delta_{qs} + \pd{N^p_{rs}}{y_q} \right) \ ,
$$
for $N_{rs} \in [H^1_{\#}(Q_1)]^d$ being the solutions to the cell problem
\begin{equation}\label{1}
\left. \begin{aligned}
-\nabla_y \cdot \left( C^{(2)} \nabla_y N_{rs} \right) = 0, \quad \text{in $Q_1$,} \\
C^{(2)}_{ijpq} \pd{N^p_{rs}}{y_q} n_j  = - C^{(2)}_{ijrs} n_j , \quad \text{on $\partial{Q_2}$.} \end{aligned} \  \right\}
\end{equation}
The tensor $\hat{C}$ is well known to be the positive homogenised elasticity tensor for a perforated elastic body with elasticity tensor $C^{(2)}$, see e.g. \cite{zhikov3}. 

For a fixed $\varphi \in [C^\infty(\Omega)]^d$, let $\varphi^0(x) \in [H^1_0(\Omega)]^d$ be the unique solution to
\begin{equation}
\label{2}
\nabla_x \cdot \Big( \hat{C} \nabla_x \varphi^0 \Big) =  \nabla_x  \cdot \varphi .
\end{equation}
Such a $\varphi^0$ exists by the positivity of $\hat{C}$. Furthermore, $\varphi^0 \in [C^\infty(\Omega)]^d$, $\varphi^0\vert_{\partial{\Omega}} = 0$. Next, take $\varphi^1 \in C^\infty(\Omega ; H^1_{\#}(Q_1))$ to be a solution to
\begin{equation}\label{3}
\left. \begin{aligned}
-\nabla_y \cdot \Big( C^{(2)} \left[ \nabla_y \varphi^1(x,y) + \nabla_x \varphi^0(x) \right]\Big) = 0, \quad \text{in $Q_1$,} \\
C^{(2)} \nabla_y \varphi^1(x,y) \cdot n  = - C^{(2)} \nabla_x \varphi^0(x) \cdot n  , \quad \text{on $\partial{Q_2}$.} \end{aligned} \  \right\}
\end{equation}
Setting
\begin{equation}
\Psi(x,y) : = \chi_1(y) \left( C^{(2)}(y) \right)^{1/2} \Big[ \nabla_y \varphi^1(x,y) + \nabla_x \varphi^0 (x) \Big] ,
\end{equation}
by construction, $\Psi \in C^\infty(\Omega ; W)$, see \eqref{spaceW}, $\Psi(x,y) = 0$ for $y \in Q_1$. Furthermore, $\varphi^1_p(x,y) = N^p_{rs} \pd{\varphi^0_r}{x_s}$, for $N_{rs}$ solving \eqref{1}, and
\begin{equation}
\label{4}
\int_{Q_1} \left( C^{(1)} (y) \right)^{1/2} \Psi(x,y) \ \mathrm{d}y = \left( \int_{Q_1} C^{(2)}_{ijpq}(y) \left( \delta_{pr}\delta_{qs} + \pd{N^p_{rs}}{y_q} \right) \ \mathrm{d}y \right)  \pd{\varphi^0_r}{x_s} = \hat{C} \nabla_x \varphi^0 . 
\end{equation}
Therefore,  \eqref{2} and \eqref{4} imply \eqref{xreq1}. It remains to prove \eqref{xreq2}. For fixed $\xi \in L^2(\Omega ; W)$
\begin{multline*}
\dblint{\Omega}{ \xi \cdot \Psi} = \int_{\Omega} \int_{Q_1}\xi \cdot \left( C^{(2)}(y) \right)^{1/2} \nabla_x \varphi^0 \ \mathrm{d}y\mathrm{d}x  \\ \le \left( \int_\Omega \left\vert \nabla_x \varphi^0 \right\vert^2  \right)^{1/2}\left( \int_\Omega \int_{Q_1} C^{(2)} \xi \cdot \xi \ \mathrm{d}y\mathrm{d}x \right)^{1/2}.
\end{multline*}
Now inequality \eqref{xreq2} follows from \eqref{sympos} and observing that
\begin{multline*}
\int_\Omega \left\vert \nabla_x \varphi^0 \right\vert^2 \le \nu \int_{\Omega} \hat{C} \nabla_x \varphi^0 \cdot \nabla_x \varphi^0 = \nu \int_\Omega \varphi \cdot \nabla_x \varphi^0 \\
\le \nu   \left( \int_\Omega \left\vert \nabla_x \varphi^0 \right\vert^2  \right)^{1/2}  \left( \int_\Omega \left\vert \varphi \right\vert^2  \right)^{1/2},
\end{multline*}
where the first inequality is due to the positivity of $\hat{C}$ and then recalling \eqref{2}.
\end{proof}
\subsection{Proof of main condition \eqref{eq:pdhom14}}
\label{sec:spaceV}
In this section we prove the validity of the key assumption \eqref{eq:pdhom14} for the tensor $C^{(1)}$ given by \eqref{elast:degentensor}-\eqref{sympos}.  That is we wish to show: there exists a constant $c>0$ such that for any $u \in [H^1_{\#}(Q)]^d$
\begin{equation}
\label{eq:elastcoercive}
\norm{P_{V^\perp} u }{H^1(Q)}^2 \le c \left( \norm{e(u)}{L^2(Q_1)}^2 + \norm{\nabla \cdot u}{L^2(Q_2)}^2 \right).
\end{equation}
Here $P_{V^\perp}$ is the projection on to $V^\perp$, the orthogonal complement of 
\begin{multline}
\label{eq:elastV1}
V = \{ v \in [H^1_{\#}(Q)]^d : v(y) = k + \chi_2(y)w(y) \text{ for some $k \in \mathbb{R}^d$, $w \in [H^1_0(Q_2)]^d$,}  \\  \nabla_y \cdot w=0 \},
\end{multline}
see \eqref{spaceV2}. The inequality \eqref{eq:elastcoercive} holds trivially for any element of $V$ and need only be validated for elements of $V^\perp$. Moreover, it is sufficient to show inequality \eqref{eq:elastcoercive} for the following equivalent $H^1_{\#}(Q)$-norm 
\begin{equation}
\label{eq:elastV2}
\norm{u}{H}^2 : = \left\vert \int_{Q_1} u \ \mathrm{d}y \right\vert^2 + \int_{Q} \vert \nabla u \vert^2 \ \mathrm{d}y,
\end{equation}
using the convention $\int u \ \mathrm{d}y = \left( \int u_1 \ \mathrm{d}y , \ldots,\int u_d \ \mathrm{d}y \right)$ for  $u = (u_1,\ldots,u_d)$. The norm \eqref{eq:elastV2} is induced by the following inner product
\begin{equation}
\label{eq:elastV3}
(u,v)_{H} : = \left(\int_{Q_1} u \ \mathrm{d}y\right) \cdot \left(\int_{Q_1} v \ \mathrm{d}y\right) + \int_{Q} \nabla u \cdot \nabla v \ \mathrm{d}y,
\end{equation}
and by definition, $w \in V^{\perp}$ if
\begin{align}
\label{eq:elastV4}
\left(\int_{Q_1} w \ \mathrm{d}y\right)\left(\int_{Q_1} v \ \mathrm{d}y\right) + \int_{Q} \nabla w \cdot \nabla v \ \mathrm{d}y = 0 \quad \forall v \in V.
\end{align}
As constant vectors are in $V$, see \eqref{eq:elastV1}, $\quad \int_{Q_1} w \ \mathrm{d}y  = 0$  and \eqref{eq:elastcoercive} easily follows from \eqref{eq:elastV2} and the following result:
\begin{lemma}
\label{lem:elastV1}
There exists a constant $c >0$ such that
\begin{equation}
\label{eq:elastV7}
\int_{Q} \vert \nabla w \vert^2 \ \mathrm{d}y \le c\left( 
\int_{Q_1} \vert e(w) \vert^2 \ \mathrm{d}y + \int_{Q_2} \vert \nabla \cdot w \vert^2 \ \mathrm{d}y \right) \quad \forall w \in V^\perp.
\end{equation}
\end{lemma}
\begin{proof}
%
%
Functions $\phi \in [C^{\infty}_0(Q_2)]^d$ with $\nabla \cdot \phi = 0$ belong to $V$, and by \eqref{eq:elastV4} $w \in V^\perp$ if
\begin{flalign}
\label{edit..1}
0 = & (w,\phi)_{H} = \int_{Q} \nabla w \cdot \nabla \phi \ \mathrm{d}y = - \mv{\Delta w , \phi}, 
\end{flalign}
where $\mv{\Delta w , \phi}$ denotes the action of the distribution $\Delta w$ on the test function $\phi$. Equation \eqref{edit..1} states the distribution $\Delta w$ is orthogonal to all divergent free test functions in $Q_2$. It is well known, see e.g. \cite[Proposition 1.1, p.~14]{temam}, that such distributions are potentials, i.e. $\Delta w = \nabla \psi$ for some distribution $\psi$ and, since $w \in [H^1(Q_2)]^d$, $\psi \in L^2(Q_2)$. Therefore, we see that $w \in V^{\perp}$ if, and only if, 
\begin{align*}
\Delta w = \nabla \psi \quad \text{in $Q_2$ for some  $\psi \in L^2(Q_2)$ }, &  \quad \int_{Q_1} w \ \mathrm{d}y  = 0.
\end{align*} 
For fixed $w \in V^{\perp}$, let $\tilde{w}$ be its harmonic extension, see Lemma \ref{lem:harmonicext} below. Denote $w_1 := w - \tilde{w}$. Evidently, $w =\tilde{w} + w_1$, $w_1 \in [H^1_{0}(Q_2)]^d$, $\int_{Q_1} w \ \mathrm{d}y = 0$,  and 
$$\begin{array}{ccc}
\tilde{w} = w \ \mathrm{in}\ Q_1, & & \Delta \tilde{w} = 0 \ \mathrm{in}\ Q_2, \\
w_1 = 0 \ \mathrm{in}\ Q_1, & & \quad \Delta w_1 = \nabla \varphi \
\mathrm{in}\ Q_2,
\end{array}$$
where $\varphi \in L^2(Q_2)$. Since
\begin{align*}
\int_{Q} \left\vert \nabla w  \right\vert^2 \ \mathrm{d}y & \le \int_{Q} \left\vert \nabla \tilde{w}  \right\vert^2 \ \mathrm{d}y + \int_{Q_2} \left\vert \nabla w_1 \right\vert^2 \ \mathrm{d}y, \\
\intertext{and}
\int_{Q_2} \left\vert \nabla \cdot w_1  \right\vert^2 & \le  \int_{Q_2} \left\vert \nabla \cdot w  \right\vert^2 + \int_{Q} \vert \nabla \tilde{w} \vert^2,
\end{align*}
to show inequality \eqref{eq:elastV7}, it is sufficient to prove the following inequalities
\begin{align}
\int_{Q} \left\vert \nabla \tilde{w}  \right\vert^2 \ \mathrm{d}y \le c
\left( \int_{Q_1} \left\vert e(w)  \right\vert^2 \ \mathrm{d}y
\right), \label{eq:elastV8} \\
\int_{Q_2} \left\vert \nabla w_1 \right\vert^2 \ \mathrm{d}y \le c \left( \int_{Q_2} \left\vert \nabla \cdot w_1  \right\vert^2 \ \mathrm{d}y
\right) \label{eq:elastV9}.
\end{align}
Inequality \eqref{eq:elastV8} directly follows from Lemma \ref{lem:harmonicext}. (iii) and Lemma \ref{lem:Kornperiodic}. 

Let us now show that inequality \eqref{eq:elastV9} holds: let $w_n \in C^\infty_0(Q_2)$ be such that $w_n \rightarrow w_1$ strongly in $H^1$. Then,  by integration by parts and Lemma \ref{lem:elastV2}
\begin{equation}
\begin{split}
\label{eq:elastV10}
\int_{Q_2} \nabla w_n \nabla w_1  \ \mathrm{d}y & = - \mv{w_n , \Delta w_1} = - \mv{w_n , \nabla\varphi}
\\ & = \int_{Q_2} \varphi \nabla\cdot w_n \ \mathrm{d}y \\ & \le \norm{\varphi}{L^2(Q_2)}
\norm{\nabla \cdot w_n}{L^2(Q_2)} \\ & \le c \norm{\nabla \cdot
w_n}{L^2(Q_2)} \left( \norm{\nabla \varphi}{H^{-1}(Q_2)} +
 \left\vert  \int_{Q_2} \varphi \ \mathrm{d}y \right\vert \ \right).
\end{split}
\end{equation}
For fixed $w \in H^1_0(Q_2)$
\begin{flalign*}
\norm{\Delta w}{H^{-1}} & = \sup_{\substack{u \in H^1_0(Q_2) \\ \norm{u}{H^1(Q_2)}=1}} \left\vert \int_{Q_2} \nabla w \cdot \nabla u  \right\vert \\ & \le \sup_{\substack{u \in H^1_0(Q_2) \\ \norm{u}{H^1(Q_2)}=1}} \norm{\nabla w}{L^2(Q_2)}\norm{ \nabla u}{L^2(Q_2)} \\
& \le \sup_{\substack{u \in H^1_0(Q_2) \\ \norm{u}{H^1(Q_2)}=1}} \norm{\nabla w}{L^2(Q_2)}\norm{u}{H^1(Q_2)} \\ 
& =  \norm{\nabla w}{L^2(Q_2)} =  \norm{ w}{H^1_0(Q_2)},
\end{flalign*}
which implies $\Delta$ defines a bounded linear operator from $H^1_0$ to $H^{-1}$. Therefore
\begin{flalign*}
\norm{\nabla \varphi}{H^{-1}(Q_2)} & = \norm{\Delta w_1}{H^{-1}(Q_2)} \le \norm{ \nabla w_1}{L^2(Q_2)}.
\end{flalign*}
Without loss of generality, because adding a constant to $\varphi$ does not affect $\nabla \varphi$,  we can choose $\int_{Q_2} \varphi \ \mathrm{d}y = 0$. Hence \eqref{eq:elastV10} becomes 
\begin{flalign*}
\int_{Q_2} \nabla w_n \nabla w_1 \ \mathrm{d}y & \le 
c\left( \int_{Q_2} \left\vert \nabla \cdot w_n \right\vert^2 \ \mathrm{d}y  \right)^{\frac{1}{2}}\left( \int_{Q_2} \left\vert \nabla w_1 \right\vert^2 \ \mathrm{d}y  \right)^{\frac{1}{2}}.
\end{flalign*}
Passing to the limit $n \rightarrow \infty$ gives the desired result:
\begin{flalign*}
\int_{Q_2} \left\vert \nabla w_1 \right\vert^2 \ \mathrm{d}y & \le 
c\left( \int_{Q_2} \left\vert \nabla \cdot w_1 \right\vert^2 \ \mathrm{d}y  \right)^{\frac{1}{2}}\left( \int_{Q_2} \left\vert \nabla w_1 \right\vert^2 \ \mathrm{d}y  \right)^{\frac{1}{2}},
\end{flalign*}
implying \eqref{eq:elastV9}.
\end{proof}
\subsection{On the technical lemmata utilised in the proof of Lemma \ref{lem:elastV1}}
\label{auxlemproofs}
\begin{lemma} ( \rm{see also Ladyzhenskaya} \cite{lady1}.)
\label{lem:elastV2}
There exists a constant $c>0$ such that for all $u \in L^2(Q_2)$
\begin{equation*}
\norm{u}{L^2(Q_2)} \le c \left(  \norm{\nabla u
  }{H^{-1}(Q_2)} + \left\vert \int_{Q_2} u\ \mathrm{d}y \right\vert \right).
\end{equation*}
\end{lemma}
\begin{proof}
Assume the contrary. Then there exists $\seq{u_n}$ in $L^2$ such that $\norm{u_n}{L^2(Q_2)}=1$ for all  $n
\in \mathbb{N}$, and 
\begin{equation}\label{eq:elastV6}
1 = \norm{u_n}{L^2(Q_2)} \ge n \left(  \norm{\nabla u_n
  }{H^{-1}(Q_2)} + \left\vert \int_{Q_2} u_n\ \mathrm{d}y \right\vert \right). \end{equation}
Since $L^2(Q_2) \hookrightarrow H^{-1}(Q_2)$ is a compact embedding, $\seq{u_n}$ has a
convergent subsequence in $H^{-1}$ to a limit $u_0$ say. After
passing to the necessary subsequence we have 
$$
u_n \longrightarrow u_0 \ \mathrm{in}\ H^{-1}.
$$ 
By Lions Lemma, c.f. \cite[Theorem 3.2, Remark 3.1, p.~111]{Ineqmechphys},
$$\norm{u_n - u_m}{L^2} \le c \left(  \norm{\nabla u_n}{H^{-1}}
 + \norm{\nabla u_m}{H^{-1}} +\norm{u_n - u_m}{H^{-1}} \right) \quad
n,m \in \mathbb{N},$$
which via \eqref{eq:elastV6} implies $\seq{u_n}$ is a Cauchy sequence in $L^2$ and, therefore, by the completeness of $L^2$, converges to a
limit $\bar{u}$ say. Since $L^2$ is embedded into $H^{-1}$ we have
$$ \norm{u_n - \bar{u}}{H^{-1}} \le c \norm{u_n - \bar{u}}{L^2}
\longrightarrow 0 \ \mathrm{as}\ n \longrightarrow \infty,$$ 
which implies $\bar{u}=u_0$, and also $\int_{Q_2} u_n \ \mathrm{d}y \longrightarrow
\int_{Q_2} u_0 \ \mathrm{d}y$. Furthermore, $\nabla u_n \longrightarrow \nabla u_0$ in
$H^{-1}$ since: for $i=1,\cdots,n$, 
\begin{align*}
\left\vert \ \mv{u_{n,i} - u_{0,i}, v} \ \right\vert & = \left\vert
  -\mv{u_n - u_0,v_{,i}}  \right\vert \\ &\le c
\norm{v_{,i}}{L^2}\norm{u_n - u_0}{L^2} \qquad \forall v \in H^1_0(Q_2).
\end{align*}
From \eqref{eq:elastV6} we see $\norm{\nabla u_n}{H^{-1}} \longrightarrow 0$,
$ \left\vert \int_{Q_2} u_n \ \mathrm{d}y \right\vert \longrightarrow 0$ as
$n \longrightarrow \infty$. Therefore
$$ \begin{matrix}
\norm{\nabla u_0}{H^{-1}} = 0, & \qquad & \int_{Q_2} u_0 \ \mathrm{d}y = 0,
\end{matrix} $$
which implies $u_0 = 0$. Hence a contradiction as, from the construction of
$\{u_n \}$, $\norm{u_0}{L^2} = 1$.
\end{proof}
\begin{lemma}
\label{lem:harmonicext}
For $u \in H^1_{\#}(Q)$ there exists $\tilde{u} \in H^1_{\#}(Q)$ and a constant $c>0$ independent of $u$ such that
\begin{enumerate}[(i)]
\item $\tilde{u} = u$ in $Q_1$,
\item $\Delta \tilde{u} = 0$ in $Q_2$,
\item $\norm{\nabla \tilde{u}}{L^2(Q)} \le c \norm{u}{H^1_{\#}(Q_1)}$. 
\end{enumerate}
We shall call $\tilde{u}$ the harmonic extension of $u$.
\end{lemma}
\begin{proof}
For fixed $u \in H^1_{\#}(Q)$, Sobolev Extension theorem, c.f. e.g. \cite{zhikov3}, says that there exists an extension operator $E: H^1_{\#}(Q_1) \rightarrow H^1_{\#}(Q)$ such that
\begin{equation}
\label{star}
\norm{Eu}{H^1(Q)} \le c \norm{u}{H^1(Q_1)},
\end{equation}
for some constant $c>0$ independent of $u$. Denote by $\tilde{u} \in H^1(Q)$ the solution to
\begin{align*}
- \Delta \tilde{u} & = 0  \quad \text{ in } Q_2, & \tilde{u} & = u  \quad \text{ on } \partial{Q_2},
\end{align*}
extended by $u$ into $Q_1$; $\tilde{u}$ satisfies (i) \& (ii). Since $\tilde{u}$ minimises the functional 
$$
F(u) = \int_{Q_2} \vert \nabla u \vert^2 \mathrm{d}y
$$
on $\{u + H^1_0(Q_2) \}$ we have, in particular,
\begin{flalign*}
\int_{Q_2} \vert \nabla \tilde{u} \vert^2 \ \mathrm{d}y \le  \int_{Q_2} \vert \nabla \left(Eu\right) \vert^2 \ \mathrm{d}y.  
\end{flalign*}
Inequality (iii) follows from \eqref{star}.
\end{proof}
\begin{lemma}
\label{lem:Kornperiodic}
Let $u \in [H^1_{\#}(Q_1)]^d$, i.e. $u \in [H^1(Q_1)]^d$ and $Q$-periodic. Then there exists a constant $c>0$ independent of $u$, such that
$$
\norm{u}{H^1(Q_1)}^2 \le c \left( \left\vert \int_{Q_1} u \ \mathrm{d}y \right\vert^{2} + \int_{Q_1} \vert e(u) \vert^2 \ \mathrm{d}y \right)
$$
\end{lemma}
\begin{proof}
As in the proof of Lemma \ref{lem:elastV2}, using the standard Korn's inequality in the place of Lions lemma c.f. \cite{Ineqmechphys}, we construct a subsequence $u_n$ that converges weakly in $H^1$ and strongly in $L^2$ to some $u_0$, $\norm{u_0}{L^2}=1$ with $\int_{Q_1} u_0 =0$ and $e(u_0) = 0$. Now, $e(u_0)=0$ implies $u_0$ is a rigid body motion but, since $u_0$ is periodic, $u_0$ is not a rotation. Therefore, $u_0$ is constant, in particular $u_0 = 0$ and the contradiction follows.
\end{proof}

\section{On the spectral convergence}
\label{sec:spectral}
In this section we shall prove the spectral compactness result, Theorem \ref{spectralcompactness}. First, we note the validity of the result is subject to the following geometric alteration: the inclusions intersecting or touching the boundary are ``excluded", i.e. we re-define $C^\ep(x)$ to equal $C^2\left( \tfrac{x}{\ep} \right)$ on such inclusions. Next, we give a precise definition as to the meaning of $A^\ep$ and $A^0$. 

Denote by $H$ the closure of $[H^1_0(\Omega)]^d \times [L^2(\Omega ; H^1_0(Q_2))]^d$ in $L^2(\Omega \times Q)$ with inner product
$$
\left( u, v \right)_H = \dblint{\Omega}{ u(x,y) \cdot v(x,y)}.
$$
Theorem \ref{thm:elastprob2} implies that 
$$
\mathcal{A}^0(u,v) : = \dblint{\Omega_1}{C^{\rm{hom}} \nabla_x u \nabla_x v}  + \dblint{\Omega_0}{ a^{(0)} \nabla_y u \cdot \nabla_y v} ,
$$
defines a bilinear form on the Hilbert space $H$. The bilinear form $\mathcal{A}^0$ is, clearly, non-negative, and is closed on $H$. Therefore, it is well known, see e.g. \cite{SiRe}, that $\mathcal{A}^0$ defines a non-negative self-adjoint operator $A^0$, called the Friedrichs extension, by
$$
\mathcal{A}^0(u,v) = (A^0 u,v)_H, \quad \forall u, v \in \mathcal{D}(A^0),
$$
where the domain $\mathcal{D}(A^0)$ is a dense subset of $H$. We call $A^0$ the homogenised limit operator. Similarly, we denote by $A^\ep$ the Friedrichs extension associated to the bilinear form \eqref{abc1}.  
\subsection{Convergence of spectra}
\label{subsec:spec}
\begin{lemma} 
\label{lem:elastspectcom1}
The spectrum of $A^\ep$, $\sigma(A^\ep)$, converges in the sense of Hausdorff to the spectrum of $A^0$, $\sigma(A^0)$. That is 
\begin{enumerate}[(i)]
\item For every $\lambda \in \sigma(A^0)$ there exists $\lambda^\ep \in \sigma(A^\ep)$ such that $\lambda^\ep \rightarrow \lambda$.
\item If there exists $\lambda^\ep \in \sigma(A^\ep)$ such that $\lambda^\ep \rightarrow \lambda$, then $\lambda \in \sigma(A^0)$.
\end{enumerate} 
\end{lemma}
According to Lemma \ref{prop:app2s}, property (i) is implied by  the strong two-scale resolvent convergence of $A^\ep$ to $A$, which was proven in Theorem \ref{thm:elastprob2}. We shall prove property (ii) by arguments that are conceptually similar to \cite{zhikov1}. Assuming $\lambda^\ep \rightarrow \lambda^0$, let $u^\ep$ be the corresponding normalised eigenfunctions of $\lambda^\ep$, i.e.
\begin{align}
\label{eq:elastspectcom1}
A^\ep u^\ep = \lambda^\ep u^\ep, & & \norm{u^\ep}{L^2(\Omega)} = 1.
\end{align}
Since the sequence $u^\ep$ is bounded, $u^\ep \twoscale u^0$ for some $u^0 \in L^2(\Omega \times Q)$. By  Theorem \eqref{thm:elastprob2}, we can pass to the limit $\ep \rightarrow 0$ and find
$$
A^0 u^0 = \lambda^0 u^0,
$$
To assure $\lambda^0$ is in the spectrum of $A^0$ is to show $u^0$ is not identically zero. 

The quadratic form
\begin{equation}
\label{abc2}
\mathcal{B}(u,v) : = \int_{Q_2} \nabla_y u(y) \cdot \nabla_y v(y) \ \mathrm{d}y
\end{equation}
on the domain $H: = \{ v \in H^1_0(Q_2) : \nabla \cdot v = 0 \}$ defines a self-adjoint operator $B$. It is well known, cf. \cite{temam}, the operator $B$ has a compact resolvent and therefore a discrete spectrum of eigenvalues going to infinity. Furthermore, by Lemma \ref{lem:elastspectcom3} $\sigma{(B)} = \{ \mu_n : n \in \mathbb{N} \} \subset \sigma{(A^0)}$. To prove $u^0 \neq 0$ it is sufficient to show the following strong two-scale compactness result 
\begin{lemma}
\label{lem:elastspectcom2}
Suppose that 
\begin{align*}
A^\ep u^\ep = \lambda^\ep u^\ep, & & \norm{u^\ep}{L^2(\Omega)} = 1.
\end{align*}
Let $\lambda^\ep \rightarrow \lambda \notin \sigma(B).$ Then $u_{\ep}$ has a strongly two-scale convergent subsequence.
\end{lemma}
Indeed, if Lemma \ref{lem:elastspectcom2} holds then, by Definition \eqref{dfn:strongtwoscale}, $\norm{u^\ep}{L^2(\Omega)} \rightarrow \norm{u^0}{L^2(\Omega \times Q)}$ which implies $u^0 \neq 0$.
\begin{proof}
Let $ \mathcal{J} : = \left\{ j \in \mathbb{Z}^d : \ep( Q + j) \cap \Omega \right\}$, and $u^\ep \vert_j$ to be the restriction of $u^\ep$ to $j$. By the geometric alteration made at the beginning of Section \ref{sec:spectral} we can use Lemma \ref{lem:moddd} below, to construct a $\hat{u}^\ep \vert_j$ for all $j \in \mathcal{J}$, and therefore construct a $\hat{u}^\ep \in H^1_0(\Omega)$ such that $\hat{u}^\ep = u^\ep$ in $\Omega^\ep_1$; $\nabla \cdot \hat{u}^\ep =\nabla \cdot u^\ep$ in $\Omega^\ep_2$ and $\Delta \hat{u}^\ep = \nabla \phi^\ep$ for some $\phi^\ep \in L^2(\Omega^{\ep}_2)$. A straightforward rescaling of Lemma \ref{lem:moddd} (iii), inequalities \eqref{elast:aprioribounds1}-\eqref{elast:aprioribounds3} and $\norm{u^\ep}{L^2(\Omega)} = 1$ show $\norm{\hat{u}^\ep}{H^1(\Omega)} \le C$. Therefore, a subsequence of $\hat{u}^\ep$ convergences strongly in $L^2(\Omega)$ to some $\hat{u}$. To prove the result it remains to show the difference $v^\ep = u^\ep - \hat{u}^\ep$ strongly two-scale converges.  

By construction $v^\ep \in [H^1_0(\Omega^\ep_2)]^d$, $\nabla_y \cdot v^\ep=0$ and since $u^\ep$ solves
\begin{align*}
\int_{\Omega} C^{(1)}\nabla u^\ep \cdot \nabla \phi \ \mathrm{d}x + \ep^2 \int_{\Omega^{\ep}_2} C^{(0)}\nabla u^\ep \cdot \nabla \phi \ \mathrm{d}x  = \int_{\Omega} \lambda^\ep u^\ep \phi \ \mathrm{d}x & \quad \forall \phi \in C^{\infty}_0(\Omega),
\end{align*}
we see that $v^\ep$ solves
\begin{align}
\label{eq:elastspectcom3}
\ep^2 \int_{\Omega^{\ep}_2} \nabla v^\ep \cdot \nabla \phi \ \mathrm{d}x = \int_{\Omega} \lambda^\ep(v^\ep + \hat{u}^\ep)\phi \ \mathrm{d}x,& & \forall \phi \in C^{\infty}_0(\Omega^{\ep}_2) \text{ such that $\nabla \cdot \phi =0.$}
\end{align}
Let us consider the following variational Stokes problem: Find $w^\ep \in [H^1_0(\Omega^\ep_2)]^d$, $\nabla_y \cdot w^\ep=0$  such that
\begin{align}
\label{eq:elastspectcom4}
\ep^2 \int_{\Omega^{\ep}_2} \nabla w^\ep \cdot \nabla \phi \ \mathrm{d}x = \int_{\Omega} \lambda^\ep w^\ep \phi + f^\ep \phi \ \mathrm{d}x,& & \forall \phi \in [C^{\infty}_0(\Omega^{\ep}_2)]^d \text{ such that $\nabla \cdot \phi =0,$}
\end{align}
for a given $f^\ep$. It remains to show the following result: If $f^\ep \twoscale f$ then $w^\ep \twoscale w$ where $w \in [L^2(\Omega ; H^1_0)]^d$ solves
\begin{align}
\label{eq:elastspectcom5}
\int_{\Omega}\int_{Q_2} \nabla_y w(x,y)\cdot \psi(x)\nabla_y \phi(y) \ \mathrm{d}y\mathrm{d}x  = \int_{\Omega}\int_{Q_2} \lambda w(x,y) \psi(x)\phi(y) \ \mathrm{d}y\mathrm{d}x  \nonumber \\ + \int_{\Omega}\int_{Q_2} f(x,y)\psi(x)\phi(y)\ \mathrm{d}y\mathrm{d}x, \quad \forall \psi \in C^{\infty}_0(\Omega), \phi \in [C^{\infty}_0(Q_2)]^d, \nabla_y \cdot \phi=0 .
\end{align}
Indeed, if this is true then
\begin{flalign*}
\lim_{\ep \rightarrow 0} \int_{\Omega} f^\ep v^\ep  \ \mathrm{d}x & = \lim_{\ep \rightarrow 0} \int_{\Omega} \ep^2 \nabla w^\ep \nabla v^\ep - \lambda^\ep w^\ep v^\ep \ \mathrm{d}x  = \lim_{\ep \rightarrow 0} \int_{\Omega} \lambda^\ep \hat{u}^\ep w^\ep
 \ \mathrm{d}x \\
& = \int_{\Omega}\int_{Q_2} \lambda \hat{u} w \ \mathrm{d}y\mathrm{d}x  =  \int_{\Omega}\int_{Q_2} \nabla_y v \cdot \nabla_y w - \lambda v w\ \mathrm{d}y\mathrm{d}x  \\ &=  \int_{\Omega}\int_{Q_2} fv \ \mathrm{d}y\mathrm{d}x ,
\end{flalign*}
where the first two equalities come from choosing test functions $\phi = v^\ep$ in \eqref{eq:elastspectcom4} and $\phi = w^\ep$ in \eqref{eq:elastspectcom3} respectively; the third equality uses the fact that $\hat{u}^\ep \strongtwoscale \hat{u}$ and finally we  use symmetry of limiting problem \eqref{eq:elastspectcom5}. By choosing $f^\ep = v^\ep$ we have shown, in particular, that
$$
\lim_{\ep \rightarrow 0} \int_{\Omega} (v^\ep(x))^2  \ \mathrm{d}x = \int_{\Omega}\int_{Q_2} v^2(x,y) \ \mathrm{d}y\mathrm{d}x.
$$ 
Hence $v^\ep \strongtwoscale v.$

To show \eqref{eq:elastspectcom5} we note that since the domain $\Omega^{\ep}_2$ consists of disjoint balls, the spectrum of $B^\ep$, the variational Stokes operator defined on the physical domain $\Omega^\ep_2$ by \eqref{eq:elastspectcom4}, coincides with the spectrum of $B$ on a single isolated inclusion as defined via \eqref{abc2} (by change of variables in \eqref{eq:elastspectcom4}). Therefore, since $\{ \lambda^\ep \}$ is a bounded sequence and $\lambda^\ep \notin \sigma(B)$ for small enough $\ep$, 
$$
\norm{(B^\ep - \lambda^\ep)^{-1}}{L^2} \le \frac{1}{\rho(\lambda^\ep , \sigma(A_0))} \le C,
$$
where $\rho$ is the distance function. Hence, since $w^\ep$ solves \eqref{eq:elastspectcom4},
$$
\norm{w^\ep}{L^2(\Omega)} = \norm{(B^\ep - \lambda^\ep)^{-1} f^\ep}{L^2(\Omega)} \le C \norm{f^\ep}{L^2(\Omega)} \le C.
$$
Furthermore
$$
 \int_{\Omega^{\ep}_2} \vert \ep \nabla w^\ep \vert^2 \ \mathrm{d}x = \int_{\Omega} \lambda^\ep (w^\ep)^2 + f^\ep w^\ep \ \mathrm{d}x \le C.
$$
By standard two-scale convergence arguments, we see that $w^\ep \twoscale w(x,y) \in [L^2(\Omega ; H)]^d$, $\ep \nabla w^\ep \twoscale \nabla_yw(x,y)$. Passing to the two-scale limit in \eqref{eq:elastspectcom4} gives \eqref{eq:elastspectcom5}.
\end{proof}
\begin{remark}
\label{rem1}
By arguments similar to those in \cite{zhikov2} one can appropriately modify the above proof to show Lemma \ref{lem:elastspectcom1} for the case $\Omega = \mathbb{R}^d$.
\end{remark}
Lemma \ref{lem:elastspectcom1} tells us in particular that for small enough $\ep$ the bottom of the spectrum of $A^\ep$ is a positive distance away from zero. Taking into consideration the Spectral Theory for self-adjoint operators we can see that a consequence of this is the following Poincar\'{e}-type inequality:
\begin{lemma}
\label{lem:epti}
There exists a constant $C>0$, independent of $\ep$, such that for all $u \in [H^1_0(\Omega)]^d$ and small enough $\ep$
\begin{equation}
\label{espti}
\norm{u}{L^2(\Omega)}^2 \le C \left( \norm{C^{(1)}(x/\ep) \nabla u}{L^2(\Omega^\ep_1)}^2 + \ep^2 \norm{C^{(0)}(x/\ep) \nabla u}{L^2(\Omega^\ep_2)}^2  \right).
\end{equation}
\end{lemma}

\begin{proof}
Since the bilinear form 
$$
\mathcal{A}^\ep(u,v) : = \int_{\Omega} \left( C^{(1)}(x/\ep) \nabla u(x) \cdot \nabla v(x) + \ep^2 C^{(0)}(x/\ep) \nabla u(x) \cdot \nabla v(x)  \right) \ \mathrm{d}x
$$
defines a non-negative quadratic form on the Hilbert space $[H^1_0(\Omega)]^d$, there exists a corresponding self-adjoint operator $A^\ep$ such that on its domain $\mathcal{D}(A^\ep)$ which is a a dense subset $\mathcal{D}(A^\ep)$ of [$H^1_0]^d$
$$
\mathcal{A}^\ep(u,v) = (A^{\ep}u,v)_{L^2(\Omega)}.
$$
By the classical Rayleigh variational principle,
\begin{flalign*}
\mathcal{A}^\ep(u,u) & = (A^\ep u , u) \ge \lambda^\ep_0 (u,u),
\end{flalign*}
where $\lambda^\ep_0$ is the smallest eigenvalue of $A^\ep$. By Lemma \ref{lem:elastspectcom1}, $\lambda^\ep_0 \rightarrow \lambda^0$ as $\ep \rightarrow 0$ where $\lambda_0>0$ is the smallest eigenvalue of the limit spectrum $\sigma(A^0)$. Therefore,
$$
\mathcal{A}^\ep(u,u) \ge c (u,u),
$$
for some $c >0$ and for all $u \in \mathcal{D}(A^\ep)$. This above inequality holds on $[H^1_0]^d$ by the fact $\mathcal{D}(A^\ep)$ is dense in $[H^1_0]^d$. Hence, \eqref{espti} holds.
\end{proof}
\begin{lemma}
\label{lem:moddd}
There exists a constant $c>0$, such that for all $u \in [H^1(Q)]^d$ there exists $\hat{u} \in [H^1(Q)]^d$, such that 
\begin{enumerate}[(i)]
\item $\hat{u} = u$ in $Q_1$,
\item $\nabla \cdot \hat{u} = \nabla \cdot u $ in $Q_2$,
\item $\norm{\hat{u}}{H^1(Q)}^2 \le c\left( \norm{u}{L^2(Q_1)} + \norm{e(u) }{L^2(Q_1)}^2 + \norm{ \nabla \cdot u}{L^2(Q_2)}^2 \right),$
\item $\Delta \hat{u} = \nabla \phi$ in $Q_2$ for some $\phi \in L^2(Q_2)$.
\end{enumerate}
\end{lemma}
\begin{proof}
Introducing the space
\begin{multline}
U = \{ v \in [H^1(Q)]^d : v(y) = k + \chi_2(y)w(y) \text{ for some $k \in \mathbb{R}^d$, $w \in [H^1_0(Q_2)]^d$,}  \\  \nabla_y \cdot w=0 \},
\end{multline}
we introduce $U^\perp$, the orthogonal complement to $U$ with respect to the following equivalent $H^1(Q)$-norm
$$
\norm{u}{H(Q)}^2 = \norm{u}{L^2(Q_1)}^2 + \norm{\nabla u}{L^2(Q)}^2.
$$
For fixed $u \in [H^1_{\#}(Q)]^d$, $u = u_1 + u_2$ for  $u_1 \in U$ and $u_2 \in U^{\perp}$. Define $v: = u_2 + k$, where $k$ is the value of $u_1$ in $Q_1$. Clearly $v$ satisfies (i) and (ii), as for $y \in Q_1$,
\begin{align*}
v(y) = u_2(y) + k = u_2(y) + u_1(y) = u(y),
\end{align*}
and, in $Q_2$,
\begin{align*}
\nabla_y \cdot v = \nabla_y \cdot u_2 = \nabla_y \cdot u_2 + \nabla_y \cdot u_1 = \nabla_y \cdot u.
\end{align*} 

It remains to show (iii). This can be seen by following the proof of Lemma \ref{lem:elastV1} and using, where appropriate, the standard Korn's inequality instead of Lemma \ref{lem:Kornperiodic}.
\end{proof}

\subsection{Spectrum of the two-scale homogenised limit operator}
\label{subsec:limspec}
Let us study the eigenvalues of the homogenised limit operator $A^0$ determined by Theorem \ref{thm:elastprob2}. As outlined in Section \ref{sec:probform} this requires studying the system
\begin{gather}
\label{eq:elastspect1} 
- \nabla \cdot \left( C^{\mathrm{hom}} \nabla u(x) \right)  = \lambda u(x) + \lambda<v>(x)  \quad \text{in $\Omega$}, \\ 
\begin{aligned} \label{eq:elastspect2}
 - \Delta_y v(x,y)   & = \lambda v(x,y) + \nabla_y p(x,y) \quad \text{in $Q_2$} \\
 \nabla_y \cdot v(x,y) & = 0 \quad \text{in $Q_2$}   \\
 v(x,y) & = 0 \quad \text{on $\partial{Q_2}$},
\end{aligned}
\end{gather}
for some unknown $p \in L^2(\Omega ; H^1(Q_2))$.

For the elasticity problem with elasticity tensor $C^{\mathrm{hom}}$, it is well known that the spectrum of the Dirichlet problem in $\Omega$ is discrete and consists of eigenvalues $\lambda^D_n$, $n=1,2,\ldots,\infty$, such that
$$
0 < \lambda^D_1 \le \lambda^D_2 \le \lambda^D_3 \le \ldots \qquad \lambda^D_n \rightarrow \infty
$$
 with associated eigenfunctions $u_n \in [H^1_0(\Omega)]^d$ such that 
$$
- \nabla \cdot \left( C^{\mathrm{hom}} \nabla u_n(x) \right)  = \lambda^D_n u_n(x)  \quad \text{in $\Omega$}.
$$
By setting $(v,p) \equiv (0,0)$ we see that $\lambda^D_n$ are in the spectrum of $A^0$ with corresponding eigenfunction $u^0_n(x,y) =u_n(x)$. 

As is also well known, for the Stokes spectral problem \eqref{eq:elastspect2}, the spectrum is also discrete and consists of $\mu_m$, $m=1,2,\ldots$, such that
$$
0 < \mu_1 \le \mu_2 \le \mu_3 \le \ldots \qquad \mu_m \rightarrow \infty
$$
and $v_m \in [H^1_0(Q_2)]^d$, $p_m \in H^1(Q_2)$ such that 
\begin{align*}
 - \Delta_y v_m(y)  &  = \mu_m v_m(y) + \nabla_y p_m(y) & \text{in $Q_2$ } \\
 \nabla_y \cdot v_m(y) &  = 0 & \text{in $Q_2$}.   
\end{align*}
If, for some $m$, $\mv{v_m}=0$ then $\mu_m$ is clearly in the spectrum of $A^0$ with corresponding eigenfunction $u^0_m(x,y) = v_m(y)$. For the eigenvalues $\mu_m$ whose corresponding eigenfunctions have non-zero mean, i.e. $\mv{v_m} \neq 0$, assuming $\mu_m \neq \lambda^D_n$ for all $n$ let $u_m(x) \in [H^1_0(\Omega)]^d$ be the solution of    
$$
- \nabla \cdot \left( C^{\mathrm{hom}} \nabla u_m(x) \right)  = \mu_m u_m(x) + \mu_m<v_m>(x)  \quad \text{in $\Omega$}.
$$
Then $\mu_m$ is in the spectrum of $A^0$ with corresponding eigenfunction $u^0_m(x,y) = u_m(x) + v_m(x,y)$. If $\mu_m = \lambda_n$ for some $n$ we have already shown above that $\mu_m$ lie in the spectrum of $A^0$. 

It remains to show that this exhausts all possible eigenvalues of the spectrum of $A^0$. That is the following result holds.
\begin{lemma}[Spectrum of the limit operator]
\label{lem:elastspectcom3}
The spectrum of the homogenised limit operator $A^0$, $\sigma{(A^0)}$, has the following representation: 
$$
\sigma{(A^0)} = \{ \lambda_n \ \vert \  n \in \mathbb{N}  \} \cup \{ \mu_m \ \vert \ m \ \in \mathbb{N} \}.
$$
Here $\lambda_n$ satisfies, for some non-trivial $u_n \in [H^1_0(\Omega)]^d$,
$$
- \nabla \cdot \left( C^{\mathrm{hom}} \nabla u_n(x) \right)  = \lambda_n u_n(x)  \quad \text{in $\Omega$},
$$
and $\mu_m$ satisfies, for some non-trivial $v_m \in [H^1_0 (Q_2)]^d$, $p_m \in H^1 (Q_2)$,
\begin{align*}
 - \Delta_y v_m(y)  &  = \mu_m v_m(y) + \nabla_y p_m(y) & \text{in $Q_2$ } \\
 \nabla_y \cdot v_m(y) &  = 0 & \text{in $Q_2$}.   
\end{align*}
\end{lemma}
\begin{proof}
For $\lambda_n, \mu_m$ given in Lemma \ref{lem:elastspectcom3} we have shown that
$$
 \{ \lambda_n \ \vert \  n \in \mathbb{N}  \} \cup \{ \mu_m \ \vert \ m \ \in \mathbb{N} \} \subset \sigma(A^0).
$$ 
To show the reverse inclusion it is sufficient to show that if $\lambda \neq \lambda_n$, $\lambda \neq \mu_m$, $\forall n$, $\forall m$ then for a given $f(x,y) \in [L^2(\Omega \times Q)]^d$ there exists a unique solution $u = u_0(x) + v(x,y)$, $u_0 \in [H^1_0 (\Omega)]^d$,  $v \in [L^2(\Omega ; H^1_0 (Q_2))]^d$, continuously depending on $f$ to
\begin{gather}
-\nabla_x \cdot \left( C^{\text{hom}} \nabla_x u(x) \right) - \lambda u(x) = \lambda <v>(x) + <f>(x), \label{eq:elastspectt1} \\
\intertext{and}
\begin{aligned} \label{eq:elastspectt2}
- \Delta_y v(x,y) - \lambda v(x,y) & = f(x,y) + \nabla_y p(x,y),\\
- \nabla_y \cdot v(x,y) & = 0, \\
\end{aligned}
\end{gather}
for some $p(x,y) \in [L^2(\Omega ; H^1(Q_2))]^d$. It is well known that if $\lambda \neq \mu_m$, $\forall m$, then there exists a unique $v$ to problem \eqref{eq:elastspectt2}. Furthermore, $g := \lambda<v> +<f> \in [L^2(\Omega)]^d$ and it is well known if $\lambda \notin \lambda_n$, $\forall n$, there exists a unique $u_0 \in [H^1_0(\Omega)]^d$ such that
$$
-\nabla_x \cdot \left( C^{\text{hom}} \nabla_x u(x)\right) - \lambda u(x) = g(x).
$$
The above construction ensures, by the boundedness of the appropriate inverse operators, the continuity of the solution in $f$.
\end{proof}

\section*{Acknowledgement(s)}
The author would like to thank I.V. Kamotski and V.P. Smyshlyaev for their invaluable comments pertaining to this work. This work was performed at the University of Bath and was sponsored by an EPSRC PhD studentship.

\bibliographystyle{gAPA}

\end{document}